\documentclass[aps,prb,reprint,superscriptaddress,floatfix,longbibliography]{revtex4-2}

\usepackage{graphicx}
\usepackage{amsmath,amssymb,amsfonts}
\usepackage{bm}
\usepackage[cmyk]{xcolor}
\usepackage[colorlinks=true,linkcolor=blue,citecolor=blue,urlcolor=blue]{hyperref}
\allowdisplaybreaks[4]

\begin{document}

\title{Exceptional-point-constrained locking of boundary-sensitive topological transitions in chiral non-Hermitian SSH-type lattices}

\author{Huimin Wang}
\affiliation{Institute of Theoretical Physics and State Key Laboratory of Quantum Optics Technologies and Devices, Shanxi University, Taiyuan 030006, China}

\author{Yanxin Liu}
\affiliation{Department of Physics, Fudan University, Shanghai 200438, China}

\author{Zhijian Li}
\email{zjli@sxu.edu.cn}
\affiliation{Institute of Theoretical Physics and State Key Laboratory of Quantum Optics Technologies and Devices, Shanxi University, Taiyuan 030006, China}

\author{Zhihao Xu}
\email{xuzhihao@sxu.edu.cn}
\affiliation{Institute of Theoretical Physics and State Key Laboratory of Quantum Optics Technologies and Devices, Shanxi University, Taiyuan 030006, China}

\begin{abstract}
Topological transitions in non-Hermitian systems are generally boundary sensitive: a point-gap winding transition under periodic boundary condition (PBC) and a non-Bloch bulk real-line-gap transition under open boundary condition (OBC) at $\mathrm{Re}(E)=0$ are governed by different spectra and therefore need not coincide. Here we show, for a class of chiral non-Hermitian Su--Schrieffer--Heeger (SSH)-type lattices, that these two criticalities can be locked by an exceptional-point-constrained (EP-constrained) parameter evolution. The key requirement is not the occurrence of isolated exceptional points, but the persistence of a zero-energy Bloch degeneracy along the entire sweep, which is generically exceptional in the non-Hermitian regime. In an analytically tractable limit of an extended non-Hermitian SSH chain, the EP-constrained manifolds and both transition boundaries are obtained in closed form, making the locking explicit. Away from this limit, numerical generalized-Brillouin-zone (GBZ) calculations confirm the correspondence for representative constrained sweeps, whereas unconstrained paths show that isolated exceptional points or Hermitian degeneracies do not enforce locking. We further verify the mechanism in a spinful four-band extension with branch-resolved GBZs, including strongly branch-imbalanced regimes. These results establish a path-dependent diagnostic principle: along EP-constrained sweeps in this SSH-type class, changes in PBC point-gap winding can indicate OBC non-Bloch bulk real-line-gap transitions and the corresponding changes in zero-energy boundary modes.
\end{abstract}

\maketitle

\section{Introduction}
Non-Hermitian systems provide a versatile setting for studying exceptional degeneracies, spectral winding, and unconventional topological phenomena in experimentally accessible quantum and wave platforms~\cite{PhysRevLett.80.5243,PhysRevLett.120.146402,RevModPhys.93.015005,ashida2020non,PhysRevLett.106.180403,PhysRevLett.129.070401,Poli2015,doi:10.1126/science.aar4005,PhysRevLett.120.113901,xiao2017observation,6gql-zgkb,lin2023topological}. In particular, the coexistence of spectral winding, exceptional points (EPs), and the non-Hermitian skin effect (NHSE) makes topological characterization intrinsically boundary sensitive~\cite{PhysRevLett.124.056802,PhysRevB.99.201103,PhysRevB.100.075403,PhysRevLett.124.086801,PhysRevX.8.031079,PhysRevLett.125.126402,PhysRevX.13.021007,PhysRevLett.116.133903,PhysRevLett.123.016805,PhysRevB.107.035424,PhysRevLett.121.026808,PhysRevLett.125.186802,PhysRevLett.123.246801,PhysRevLett.128.157601}. Point-gap topology and spectral winding are naturally defined from the Bloch spectrum under periodic boundary conditions (PBCs)~\cite{PhysRevLett.132.136401}. By contrast, in the presence of the NHSE, open-boundary bulk topology must be formulated in terms of non-Bloch spectra on generalized Brillouin zones (GBZs)~\cite{PhysRevX.9.041015,PhysRevLett.125.226402}. Consequently, Bloch-band information alone generally does not determine open-boundary bulk topological transitions or the associated boundary modes.

A direct consequence of this boundary sensitivity is that topological transitions identified under PBCs and OBCs need not coincide. For a fixed reference energy, a PBC point-gap winding transition occurs when the Bloch spectral loop crosses the reference point and the corresponding winding number changes. Under OBCs, the relevant bulk spectrum is instead the non-Bloch continuum defined on the GBZ. In the chiral systems considered below, the open-boundary bulk topology is characterized by a real line gap of this non-Bloch continuum at $\mathrm{Re}(E)=0$. The corresponding OBC transition occurs when the non-Bloch bulk real line gap closes, allowing the associated non-Bloch line-gap invariant to change. Throughout this work, this line gap refers to the non-Bloch bulk continuum; boundary modes, when present in a finite open chain, may lie inside this bulk gap. Since the PBC point-gap winding and the OBC non-Bloch bulk real line gap are defined from different spectra and different gap structures, their critical points are generically distinct~\cite{PhysRevResearch.7.023233,PhysRevLett.133.266604,PhysRevLett.130.203605,PRXQuantum.4.030315,liu2023simultaneous,PhysRevB.107.155430}. From an experimental perspective, directly reconstructing non-Bloch topology is often challenging~\cite{helbig2020generalized,Xiao2020,doi:10.1126/science.aaz8727,PhysRevLett.126.230402}. It is therefore useful to identify constrained situations in which PBC spectral winding can serve as a diagnostic of OBC non-Bloch bulk transitions.

This raises a central question: under what conditions can a change in PBC point-gap winding diagnose an OBC non-Bloch bulk real-line-gap transition? Along a generic parameter sweep, a change of Bloch point-gap winding may occur while the OBC non-Bloch bulk real line gap remains open, whereas an OBC non-Bloch bulk real-line-gap closing may occur without a corresponding change in the PBC winding number. A diagnostic principle is therefore needed to distinguish genuine locking of these boundary-sensitive criticalities from accidental coincidence~\cite{liu2023simultaneous}. In the chiral models considered here, the band touchings relevant to the real-line-gap transition at $\mathrm{Re}(E)=0$ are pinned to zero energy by chiral symmetry and are generically exceptional away from Hermitian or symmetry-enhanced limits. This motivates the notion of an EP-constrained path: a parameter evolution along which a zero-energy Bloch degeneracy persists throughout the sweep. The term EP-constrained emphasizes that this persistent degeneracy is generically exceptional in the non-Hermitian regime, although it may become nondefective at special Hermitian points. As shown below for Su--Schrieffer--Heeger (SSH)-type chiral lattices, this persistent zero-energy Bloch-degeneracy constraint, rather than isolated EPs, provides a path-dependent mechanism for locking the monitored PBC point-gap winding transition to the OBC non-Bloch bulk real-line-gap transition.

In this work, we demonstrate this EP-constrained locking mechanism in a class of chiral SSH-type non-Hermitian lattices~\cite{heiss2004exceptional,miri2019exceptional,chen2020revealing,PhysRevE.69.056216,doppler2016dynamically}. We first analyze an extended non-Hermitian SSH chain~\cite{PhysRevLett.42.1698,PhysRevResearch.1.023013,PhysRevB.109.155137,PhysRevA.101.013635}. In the analytically tractable limit, the EP-constrained manifolds, the PBC point-gap winding boundary, and the OBC non-Bloch bulk real-line-gap boundary can all be obtained in closed form, making the locking explicit. We then verify the same correspondence for representative constrained sweeps away from this solvable limit using numerical GBZ calculations. In contrast, unconstrained paths show that the PBC point-gap winding transition and the OBC non-Bloch bulk real-line-gap transition generally decouple, even if isolated EPs or Hermitian degeneracies are encountered. Finally, we test the mechanism in a spinful four-band extension with branch-resolved GBZs~\cite{PhysRevB.103.125411,PhysRevResearch.4.013243,PhysRevB.111.115415}, including strongly branch-imbalanced regimes. These results establish EP-constrained evolution as a path-dependent diagnostic principle for relating PBC point-gap winding transitions to OBC non-Bloch bulk real-line-gap transitions in chiral SSH-type non-Hermitian lattices.

\section{Extended non-Hermitian SSH chain and locking criterion}

In this section, we introduce a minimal chiral non-Hermitian SSH-type lattice in which PBC point-gap winding transitions and OBC non-Bloch bulk real-line-gap transitions can be directly compared. After presenting the model and its chiral symmetry, we review the corresponding boundary-dependent topological quantities: the Bloch spectral winding under PBCs and the non-Bloch real-line-gap invariant under OBCs. We then formulate the persistent zero-energy Bloch-degeneracy constraint that defines an EP-constrained path and provides the basis for the path-dependent locking mechanism analyzed in the following sections.

\subsection{Model and chiral symmetry}

\begin{figure}[htbp]
	\centering
	\includegraphics[width=\columnwidth]{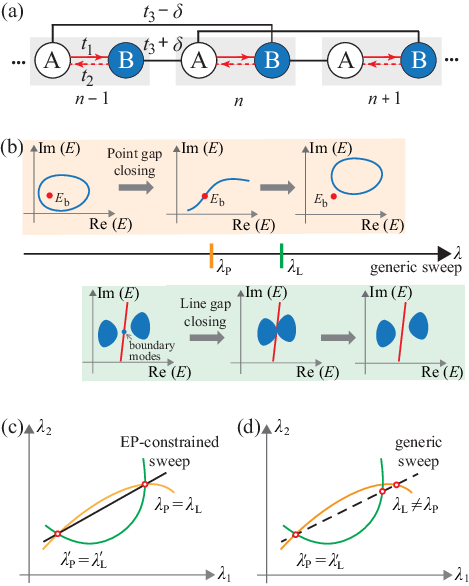}
	\caption{Boundary-sensitive topological transitions and EP-constrained locking. 
(a) Extended non-Hermitian SSH chain with nonreciprocal intracell hoppings $t_1$ and $t_2$ and reciprocal staggered intercell hoppings $t_3\pm\delta$. 
(b) Generic mismatch between a PBC point-gap winding transition at $\lambda=\lambda_{\rm P}$ and an OBC non-Bloch bulk line-gap transition at $\lambda=\lambda_{\rm L}$. In general, $\lambda_{\rm P}\neq\lambda_{\rm L}$. 
(c) Along an EP-constrained sweep, the Bloch spectrum remains pinned to a zero-energy degeneracy, and the two transition loci can be crossed at the same parameter value in the SSH-type models studied here. 
(d) Along a generic unconstrained sweep, the two loci are crossed at different parameter values, and the locking is lost.}\label{fig1}
\end{figure}

We consider a one-dimensional extended non-Hermitian SSH chain with nonreciprocal intracell hopping and staggered reciprocal intercell hopping, as illustrated in Fig.~\ref{fig1}(a). This model provides a minimal setting in which PBC point-gap winding transitions and OBC non-Bloch bulk real-line-gap transitions can be directly compared in the presence of the non-Hermitian skin effect.

The real-space Hamiltonian of the two-band model is
\begin{equation}
	\begin{aligned}
	H = \sum_{n} \Big[&
	t_1\, b_{n}^{\dagger}a_{n}
	+t_2\, a_{n}^{\dagger}b_{n} \\
	&+\left(t_3+\delta\right)
	\left(b_{n}^{\dagger}a_{n+1}+\mathrm{h.c.}\right) \\
	&+\left(t_3-\delta\right)
	\left(a_{n}^{\dagger}b_{n+1}+\mathrm{h.c.}\right)
	\Big].
	\end{aligned}
	\label{eq:real_space_hamiltonian}
\end{equation}
Here $a_{n}^{\dagger}$ ($a_{n}$) and $b_{n}^{\dagger}$ ($b_{n}$) create (annihilate) a particle on sublattices $A$ and $B$ in the $n$th unit cell, respectively. The asymmetric intracell hoppings $t_1$ and $t_2$ introduce nonreciprocity and render the Hamiltonian non-Hermitian when $t_1\neq t_2$. By contrast, the intercell hoppings are reciprocal within each channel, with amplitudes $t_3+\delta$ and $t_3-\delta$. The parameter $\delta$ therefore controls the staggering between the two inequivalent intercell hopping channels. Throughout this work, all hopping amplitudes are taken to be real.

Under PBCs, Fourier transformation gives the Bloch Hamiltonian
\begin{equation}
	H(k)=
	\begin{bmatrix}
		0 & h_2(k)\\
		h_1(k) & 0
	\end{bmatrix},
	\label{eq:bloch_hamiltonian}
\end{equation}
with
\begin{align}
	h_1(k)&=t_1+2\bigl(t_3\cos k+i\delta\sin k\bigr),\notag\\
	h_2(k)&=t_2+2\bigl(t_3\cos k-i\delta\sin k\bigr).
	\label{eq:bloch_offdiagonal}
\end{align}
The Hamiltonian is purely off diagonal and therefore obeys chiral symmetry, $\sigma_z H(k)\sigma_z^{-1}=-H(k)$. Consequently, the Bloch spectrum is symmetric about zero energy,
\begin{equation}
	E_{\pm}(k)=\pm\sqrt{h_1(k)h_2(k)}.
\end{equation}
Any Bloch band degeneracy, when present, is therefore pinned to zero energy. This zero-energy degeneracy provides the constraint used to define the EP-constrained paths below. Importantly, this pinned zero-energy degeneracy is not used as the reference point for the PBC point-gap winding along such constrained sweeps. Instead, it acts as a constraint on the Bloch spectrum, while the PBC winding is evaluated with respect to a fixed reference energy chosen inside the monitored point-gap region.
	
\subsection{PBC point-gap winding and OBC non-Bloch bulk topology}
\label{sec:pbc_obc_topology}

We next summarize the boundary-sensitive topological quantities used in this work. Non-Hermitian band spectra admit two distinct gap notions: point gaps and line gaps~\cite{RevModPhys.93.015005,PhysRevX.9.041015}. As illustrated schematically in Fig.~\ref{fig1}(b), a point gap with respect to a reference energy $E_{\rm b}$ is open when the complex-energy spectrum avoids the reference point $E_{\rm b}$ in the complex-energy plane. A line gap, by contrast, is open when the spectrum avoids a reference line, so that the bands can be continuously separated on opposite sides of that line. Correspondingly, a point-gap transition occurs when the complex spectrum crosses the reference point and the associated winding becomes ill defined, whereas a line-gap transition occurs when the spectrum touches the reference line and the corresponding line-gap invariant can change. Since these two gap-closing conditions are generally inequivalent, point-gap and line-gap transitions need not coincide in non-Hermitian systems.

In the present chiral problem, we compare two boundary-dependent criticalities: the PBC point-gap winding transition around a fixed reference energy $E_{\rm b}$ and the OBC non-Bloch bulk real-line-gap transition at $\mathrm{Re}(E)=0$. The central question is when these two otherwise distinct criticalities can become locked along a constrained parameter path.

Under PBCs, the relevant quantity is the point-gap winding number defined from the Bloch spectrum. For a chosen reference energy $E_{\rm b}$, a point gap is open when
\begin{equation}
	\det[H(k)-E_{\rm b}]\neq 0
\end{equation}
for all momenta $k$. The associated point-gap winding number is
\begin{equation}
	\nu_{\rm PBC}
	=-\frac{1}{2\pi i}\int_{-\pi}^{\pi} dk\,\partial_k \ln \det\left[H(k)-E_{\rm b}\right],
	\label{eq:pbc_winding}
\end{equation}
which measures the winding of the Bloch spectral loop around $E_{\rm b}$ as $k$ traverses the Brillouin zone. In the two-band model considered here, a nonzero $\nu_{\rm PBC}$ is associated with the NHSE in the monitored point-gap sector, and its sign tracks the dominant skin-localization direction. In particular, $\nu_{\rm PBC}=-1$ and $\nu_{\rm PBC}=+1$ correspond to opposite skin-accumulation directions along the representative sweeps studied below, whereas $\nu_{\rm PBC}=0$ indicates the absence of a net spectral winding around the chosen reference energy.

In all numerical evaluations below, $E_{\rm b}$ is kept fixed along a given parameter sweep and is chosen inside the point-gap region whose winding is monitored. Since the EP-constrained sweeps contain a pinned zero-energy Bloch degeneracy, $E_{\rm b}$ is not taken to be zero on such sweeps. Thus, when we refer to a PBC point-gap winding transition, we mean a change of the spectral winding around this chosen reference energy. The zero-energy degeneracy itself does not define an open point gap at $E_{\rm b}=0$ along the constrained path; rather, it serves as the spectral constraint imposed on the Bloch Hamiltonian.

Under OBCs, by contrast, the NHSE requires a non-Bloch description. The Bloch factor $e^{ik}$ is replaced by a complex variable $\beta$, and the open-boundary bulk continuum is obtained by evaluating $H(\beta)$ on the GBZ rather than on the unit circle. In the present chiral setting, the relevant line gap refers to the non-Bloch bulk spectrum: the bulk continuum is required to avoid the reference line $\mathrm{Re}(E)=0$. Boundary modes, when present in a finite open chain, may lie inside this bulk line gap. Thus, throughout this work, an OBC non-Bloch bulk real-line-gap transition means a closing of the non-Bloch bulk real line gap, not the absence of in-gap boundary eigenvalues in the full finite-chain OBC spectrum.

For later use, we quantify the non-Bloch bulk real line gap by
\begin{equation}
	\Delta_{\rm L}	=	\min_{\beta\in\mathrm{GBZ},\,n}\left|\mathrm{Re}\,E_n(\beta)\right|,
	\label{eq:line_gap_indicator}
\end{equation}
where $E_n(\beta)$ denotes the $n$th non-Bloch bulk eigenenergy evaluated on the GBZ. In the multiband cases with branch-resolved GBZs discussed below, the minimum is understood to be taken over all relevant branches and their corresponding non-Bloch bulk spectra. An open bulk real line gap corresponds to $\Delta_{\rm L}>0$, whereas $\Delta_{\rm L}=0$ marks a non-Bloch bulk line-gap closing. The line-gap invariant discussed below is meaningful only when $\Delta_{\rm L}>0$.

When the non-Bloch bulk real line gap is open, we characterize the corresponding phase by the non-Bloch invariant
\begin{equation}
	\nu_{\rm OBC}=\frac{\varphi_B}{2\pi},\qquad
	\varphi_B=\varphi_{Z+}+\varphi_{Z-},
	\label{eq:obc_invariant}
\end{equation}
where the biorthogonal Berry phases are accumulated along the GBZ, parameterized as $\beta=\beta(\theta)$:
\begin{equation}
	\varphi_{Z\pm}
	=-\oint d\theta\,
	\frac{\langle \psi^{(L)}_{\pm}(\theta)|\, i\,\partial_{\theta}\,|\psi^{(R)}_{\pm}(\theta)\rangle}
	{\langle \psi^{(L)}_{\pm}(\theta)|\psi^{(R)}_{\pm}(\theta)\rangle}.
	\label{eq:biorthogonal_berry_phase}
\end{equation}
Here $|\psi^{(R)}_{\pm}(\theta)\rangle$ and $|\psi^{(L)}_{\pm}(\theta)\rangle$ are the right and left non-Bloch eigenstates of $H[\beta(\theta)]$, respectively. The two bands remain paired by chiral symmetry. The invariant $\nu_{\rm OBC}$ can change only when the non-Bloch bulk continuum touches $\mathrm{Re}(E)=0$, i.e., when $\Delta_{\rm L}$ vanishes~\cite{PhysRevLett.121.086803,PhysRevLett.121.026808}. In the present two-band setting, this non-Bloch bulk invariant determines the appearance or disappearance of zero-energy boundary modes in finite open chains. The explicit non-Bloch Hamiltonian, GBZ equal-modulus condition, and numerical construction are given in Appendix~\ref{app:two_band}.

\subsection{EP-constrained paths and locking criterion}
\label{sec:ep_constraint}

We now use the chiral structure of Eq.~\eqref{eq:bloch_hamiltonian} to formulate the degeneracy constraint that underlies the locking mechanism. A zero-energy Bloch degeneracy occurs at momentum $k_0$ when
\begin{equation}
	\det[H(k_0)]=h_1(k_0)h_2(k_0)=0.
	\label{eq:zero_degeneracy}
\end{equation}
Because the Hamiltonian is purely off diagonal, a generic solution of Eq.~\eqref{eq:zero_degeneracy} has only one vanishing off-diagonal element, either $h_1(k_0)=0$ or $h_2(k_0)=0$. In this case, $H(k_0)$ has rank one and is defective, so the zero-energy degeneracy is a second-order EP. A nondefective zero-energy degeneracy requires the additional fine tuning $h_1(k_0)=h_2(k_0)=0$, for which the Hamiltonian at $k_0$ becomes the zero matrix. Thus, away from special Hermitian or symmetry-enhanced limits, zero-energy Bloch degeneracies are generically exceptional in the present model.

This observation motivates the definition of an EP-constrained manifold. We define it as a parameter submanifold $\mathcal{M}$ in the control-parameter space such that, for every parameter point $\lambda\in\mathcal{M}$, there exists at least one momentum $k_0$ satisfying
\begin{equation}
	\det[H(k_0;\lambda)]=0.
	\label{eq:ep_manifold}
\end{equation}
The momentum $k_0$ need not remain fixed along $\mathcal{M}$; rather, the defining requirement is that the Bloch spectrum remains pinned to a zero-energy degeneracy somewhere in the Brillouin zone throughout the parameter evolution. At generic non-Hermitian points on $\mathcal{M}$, this degeneracy is exceptional, whereas at fine-tuned Hermitian or symmetry-enhanced points it may become an ordinary nondefective band touching. A one-parameter sweep confined to $\mathcal{M}$ will be referred to as an EP-constrained path.

With the PBC point-gap and OBC non-Bloch bulk real-line-gap notions defined in Sec.~\ref{sec:pbc_obc_topology}, Figs.~\ref{fig1}(c) and \ref{fig1}(d) schematically illustrate the role of the EP constraint. In Fig.~\ref{fig1}(c), the parameter sweep is confined to an EP-constrained manifold, so that the Bloch spectrum remains pinned to a zero-energy degeneracy throughout the evolution. Within the chiral SSH-type models analyzed below, this persistent zero-energy degeneracy constraint provides a path-dependent mechanism that synchronizes the monitored PBC point-gap winding transition with the OBC non-Bloch bulk real-line-gap transition along the constrained sweeps considered here. In this sense, the EP-constrained manifold serves as a diagnostic locking criterion in the present model class.

By contrast, Fig.~\ref{fig1}(d) represents a generic sweep away from the constrained manifold. In that case, the PBC point-gap winding-transition locus and the OBC non-Bloch bulk real-line-gap-transition locus are generally crossed at different parameter values, and the locking is lost. This distinction emphasizes that the relevant ingredient is not the mere encounter with isolated EPs, but the persistence of the zero-energy Bloch-degeneracy constraint along the entire path.

We establish this locking explicitly below, first in an analytically tractable limit of the two-band model and then in the generic non-Bloch regime using GBZ calculations, before testing its robustness in a spinful four-band extension.

\section{Analytic locking in the solvable limit}

We first consider the analytically tractable limit $\delta=t_3$, where one of the two staggered intercell hopping channels vanishes. In this limit, the EP-constrained manifolds, the PBC point-gap winding-transition boundary, and the OBC non-Bloch bulk real-line-gap-transition boundary can all be obtained in closed form. This makes the path-dependent locking mechanism particularly transparent.

\begin{figure}[htbp]
  \centering
  \includegraphics[width=\columnwidth]{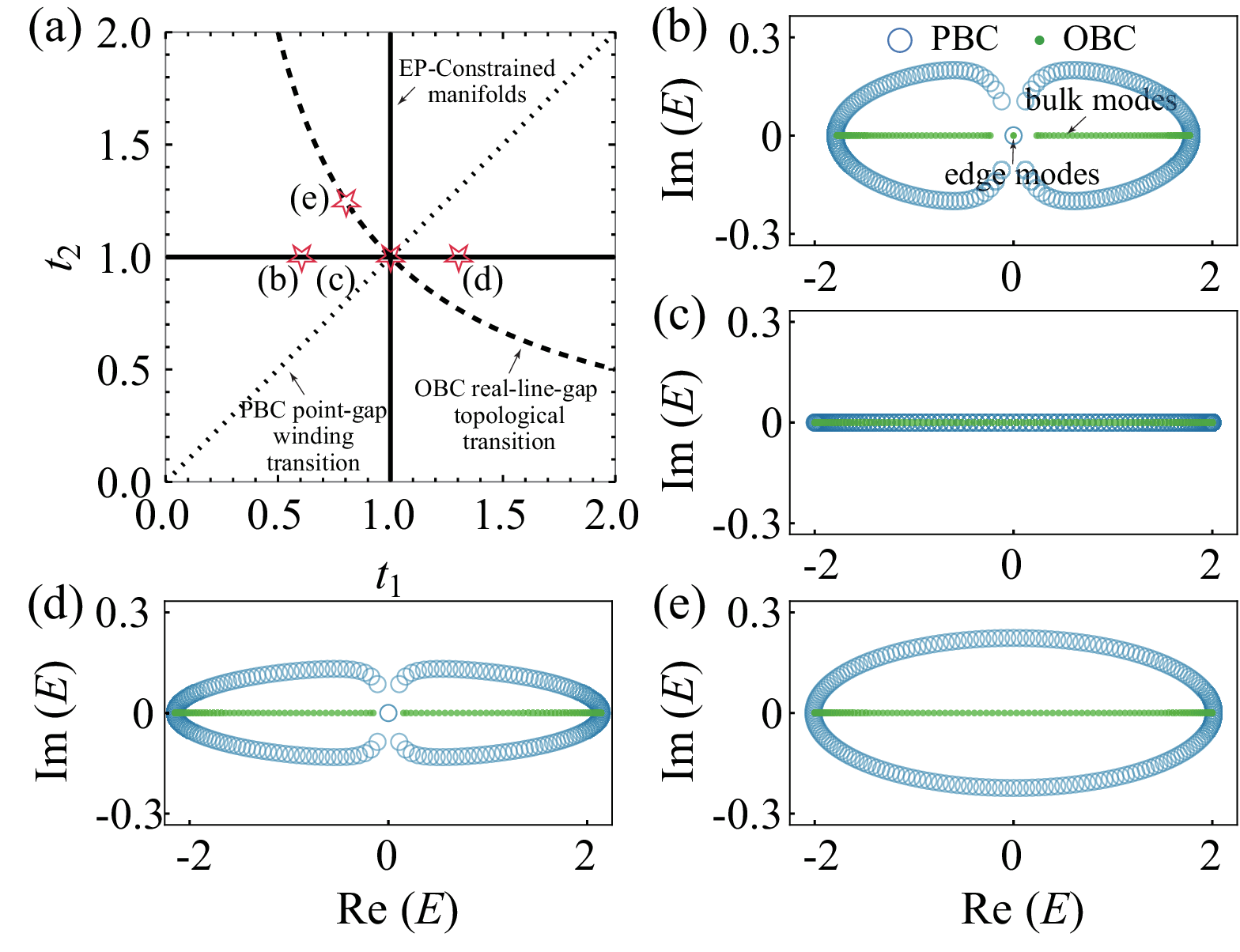}
  \caption{Analytically solvable limit $\delta=t_3=0.5$. 
  (a) Phase diagram in the $(t_1,t_2)$ plane. The solid lines are the EP-constrained manifolds, $|t_1|=2|t_3|$ or $|t_2|=2|t_3|$. The dotted line marks the PBC point-gap winding-transition boundary, $|t_1|=|t_2|$, and the dashed line marks the OBC non-Bloch bulk real-line-gap-transition boundary, $|t_3|=\sqrt{|t_1t_2|}/2$. Stars indicate the representative parameter points used in (b)--(e). 
  (b)--(e) Complex-energy spectra under PBCs (open circles) and OBCs (solid circles) with system size $N=50$ for 
  (b) $(t_1,t_2)=(0.6,1.0)$, 
  (c) $(t_1,t_2)=(1.0,1.0)$, 
  (d) $(t_1,t_2)=(1.3,1.0)$, and 
  (e) $(t_1,t_2)=(0.8,1.25)$.}
  \label{fig2}
\end{figure}

For $\delta=t_3$, the Bloch off-diagonal elements reduce to
\begin{align}
	h_1(k)&=t_1+2t_3e^{ik},\notag\\
	h_2(k)&=t_2+2t_3e^{-ik}.
	\label{eq:solvable_offdiagonal}
\end{align}
The zero-energy degeneracy condition in Eq.~\eqref{eq:zero_degeneracy} then requires either $h_1(k_0)=0$ or $h_2(k_0)=0$ for some real momentum $k_0$. This gives the EP-constrained manifolds
\begin{equation}
	|t_1|=2|t_3|
	\qquad \text{or} \qquad
	|t_2|=2|t_3|,
	\label{eq:solvable_ep_manifolds}
\end{equation}
shown as the solid lines in Fig.~\ref{fig2}(a). Along these manifolds, the Bloch spectrum remains pinned to a zero-energy degeneracy somewhere in the Brillouin zone. Away from special Hermitian points, this degeneracy is an EP.

Under OBCs, after clearing the negative power of $\beta$, the characteristic equation $\det[H(\beta)-E]=0$ becomes quadratic in $\beta$. The GBZ is therefore a circle parameterized by $\beta(\theta)=re^{i\theta}$, with radius $r=\sqrt{\left|t_1/t_2\right|}$. The corresponding skin exponent is
\begin{equation}
	\kappa=\ln r=\frac{1}{2}\ln\left|\frac{t_1}{t_2}\right|.
	\label{eq:skin_exponent_solvable}
\end{equation}
Thus the non-Bloch skin exponent changes sign when $|t_1|=|t_2|$. For the monitored point-gap sector, this condition coincides with the PBC point-gap winding-transition boundary, shown as the dotted line in Fig.~\ref{fig2}(a). Across this boundary, the point-gap winding number $\nu_{\rm PBC}$ changes sign, which corresponds to a reversal of the dominant skin-accumulation direction in this two-band setting.

The OBC non-Bloch bulk real-line gap closes when the non-Bloch bulk continuum
touches the reference line $\mathrm{Re}(E)=0$. In the present solvable limit and
real-parameter regime, this closing occurs when a zero-energy bulk solution lies
on the GBZ. At $E=0$, the condition $h_1(\beta)h_2(\beta)=0$ gives
$\beta=-t_1/(2t_3)$ or $\beta=-2t_3/t_2$. Requiring either root to satisfy
$|\beta|=r$, yields
\begin{equation}
	|t_3|=\frac{1}{2}\sqrt{|t_1t_2|},
	\label{eq:obc_transition_solvable}
\end{equation}
shown as the dashed line in Fig.~\ref{fig2}(a). Across this boundary, the non-Bloch bulk real-line-gap invariant $\nu_{\rm OBC}$ changes, accompanied by the appearance or disappearance of zero-energy boundary modes in finite open chains.

Figure~\ref{fig2}(a) makes the locking mechanism geometrically transparent. In the $(t_1,t_2)$ plane, the EP-constrained manifolds, the PBC point-gap winding-transition boundary, and the OBC non-Bloch bulk real-line-gap-transition boundary are generally distinct loci. Therefore, without an additional constraint on the parameter sweep, the PBC and OBC criticalities are not expected to coincide. However, when the sweep is restricted to an EP-constrained line, the intersections with the PBC winding-transition boundary and the OBC non-Bloch bulk real-line-gap-transition boundary occur at the same parameter value.

This locking is illustrated in Figs.~\ref{fig2}(b)--\ref{fig2}(d), taken along the horizontal EP-constrained line $t_2=2t_3=1$. For $t_1=0.6$ [Fig.~\ref{fig2}(b)], the PBC point-gap winding number is $\nu_{\rm PBC}=-1$, corresponding to left-directed skin accumulation. Meanwhile, the non-Bloch bulk continuum remains real-line gapped, and the finite open chain supports isolated zero-energy boundary modes, with $\nu_{\rm OBC}=1$. At $t_1=1.0$ [Fig.~\ref{fig2}(c)], the sweep reaches the Hermitian point $t_1=t_2$, where the zero-energy exceptional degeneracy continuously reduces to an ordinary nondefective band touching. At the same parameter value, the non-Bloch bulk continuum touches $\mathrm{Re}(E)=0$, so that the OBC bulk real line gap closes. Thus the PBC point-gap winding transition and the OBC non-Bloch bulk real-line-gap transition occur simultaneously. For $t_1=1.3$ [Fig.~\ref{fig2}(d)], the non-Bloch bulk line gap reopens, the zero-energy boundary modes disappear under OBCs, and the OBC invariant changes to $\nu_{\rm OBC}=0$. At the same time, the PBC winding changes to $\nu_{\rm PBC}=+1$, indicating right-directed skin accumulation. Hence, along this EP-constrained line, the PBC point-gap winding transition is locked to the OBC non-Bloch bulk real-line-gap transition.

By contrast, Fig.~\ref{fig2}(e) shows that this locking is not generic. The chosen point lies on the OBC non-Bloch bulk real-line-gap-transition boundary but away from the EP-constrained manifolds. In this case, the non-Bloch bulk continuum reaches the real-line-gap closing condition, whereas the corresponding PBC winding-transition condition is not satisfied at this parameter point. The OBC non-Bloch bulk real-line-gap transition therefore occurs without a concomitant PBC point-gap winding transition. This contrast demonstrates that the locking is not caused by the mere presence of a line-gap closing, but by restricting the parameter evolution to an EP-constrained manifold.

\section{Beyond the solvable limit}
\subsection{EP-constrained locking}

\begin{figure*}[htbp]
  \centering
  \includegraphics[width=0.9\textwidth]{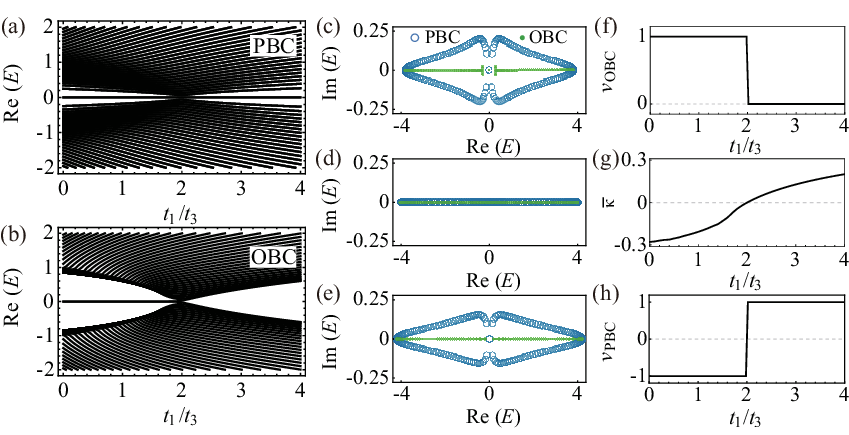}
  \caption{Locking of the OBC non-Bloch bulk real-line-gap transition and the PBC point-gap winding transition away from the solvable limit. Here $\delta=0.5t_3$ and $t_2=2t_3$, so that the PBC spectrum remains pinned to a zero-energy degeneracy throughout the sweep. 
  (a) Real parts of the PBC spectrum versus $t_1/t_3$, showing the persistent zero-energy degeneracy and the band touching at $t_1=2t_3$, where the sweep passes through the Hermitian point. 
  (b) Real parts of the finite-chain OBC spectrum. The corresponding non-Bloch bulk real line gap closes and reopens at the same point, while zero-energy boundary modes lie inside the bulk gap for $t_1<2t_3$ and disappear for $t_1>2t_3$.
  (c)--(e) Complex spectra at $t_1=1.6t_3$, $2.0t_3$, and $2.4t_3$, respectively. The PBC point-gap winding number changes from $\nu_{\rm PBC}=-1$ to $\nu_{\rm PBC}=+1$ across $t_1=2t_3$, while the OBC non-Bloch bulk real-line-gap transition and the disappearance of zero-energy boundary modes occur at the same critical point. 
  (f)--(h) Evolution of the OBC non-Bloch bulk real-line-gap invariant $\nu_{\rm OBC}$, the average skin exponent $\bar{\kappa}$, and the PBC point-gap winding number $\nu_{\rm PBC}$, respectively, showing a common critical point at $t_1=2t_3$.}
  \label{fig3}
\end{figure*}

We next move beyond the analytically tractable limit and consider the generic case $\delta\neq t_3$, where both staggered intercell hopping channels remain finite and the GBZ is no longer a simple circle. Although the closed-form simplifications of the solvable limit are lost, the same path-dependent mechanism persists within the present chiral SSH-type family: when the parameter sweep is confined to an EP-constrained manifold, the monitored PBC point-gap winding transition remains locked to the OBC non-Bloch bulk real-line-gap transition.

Figure~\ref{fig3} shows a representative EP-constrained sweep for $\delta=0.5t_3$ and $t_2=2t_3$. With this choice, $h_2(\pi)=t_2-2t_3=0$ for all values of $t_1/t_3$. Therefore, the Bloch spectrum contains a zero-energy degeneracy throughout the entire sweep. Away from the Hermitian point $t_1=t_2=2t_3$, this degeneracy is exceptional. It is important to distinguish this persistent zero-energy degeneracy from the topological transition itself. The degeneracy is the spectral constraint imposed along the path, whereas the transition occurs only at the parameter value where the PBC point-gap winding changes and the OBC non-Bloch bulk continuum simultaneously closes its real line gap. In the present sweep, this occurs at $t_1=2t_3$.

Figures~\ref{fig3}(a) and \ref{fig3}(b) display the real parts of the PBC
spectrum and the finite-chain OBC spectrum, respectively, as functions of
$t_1/t_3$. Under PBCs, the Bloch spectrum contains a zero-energy degeneracy
throughout the sweep, reflecting the imposed EP constraint. At $t_1=2t_3$, the
sweep passes through the Hermitian point, where the monitored Bloch spectral
loop crosses the fixed reference energy $E_{\rm b}$ and the corresponding point
gap closes. This point therefore marks the PBC point-gap winding transition.
Under OBCs, the relevant bulk transition is identified from the non-Bloch bulk
continuum: its real line gap closes at the same parameter value, $t_1=2t_3$,
and reopens on the other side. For $t_1<2t_3$, the non-Bloch bulk continuum is
real-line gapped, and the finite open chain supports zero-energy boundary modes
inside this bulk gap. For $t_1>2t_3$, the bulk real line gap remains open, but
the zero-energy boundary modes are absent. Thus, the monitored PBC point-gap
winding transition, the OBC non-Bloch bulk real-line-gap transition, and the
associated boundary-mode transition occur simultaneously along this
EP-constrained sweep.

The same correspondence is seen more directly in the complex spectra in Figs.~\ref{fig3}(c)--\ref{fig3}(e), corresponding to $t_1=1.6t_3$, $2.0t_3$, and $2.4t_3$, respectively. At $t_1=1.6t_3$ [Fig.~\ref{fig3}(c)], the PBC spectrum already contains a zero-energy exceptional degeneracy, while the point-gap winding number is $\nu_{\rm PBC}=-1$, corresponding to left-directed skin accumulation under OBCs. At the same parameter value, the non-Bloch bulk continuum is real-line gapped, and the finite open chain hosts zero-energy boundary modes. At the critical point $t_1=2.0t_3$ [Fig.~\ref{fig3}(d)], the sweep passes through the Hermitian point. The zero-energy exceptional degeneracy in the PBC spectrum continuously reduces to an ordinary nondefective band touching. Simultaneously, the OBC non-Bloch bulk continuum touches $\mathrm{Re}(E)=0$ and, in this case, reaches $E=0$, so that the bulk real line gap closes at the same parameter value. For $t_1=2.4t_3$ [Fig.~\ref{fig3}(e)], the PBC spectrum still contains a zero-energy degeneracy, but the monitored point-gap winding has changed to $\nu_{\rm PBC}=+1$, indicating right-directed skin accumulation. Correspondingly, the OBC non-Bloch bulk real line gap reopens, and the zero-energy boundary modes are absent in the finite open chain. Hence, although the zero-energy Bloch degeneracy persists throughout the EP-constrained sweep, the PBC point-gap winding transition occurs precisely at the OBC non-Bloch bulk real-line-gap transition.

This synchronization is further quantified in Figs.~\ref{fig3}(f)--\ref{fig3}(h). At $t_1=2t_3$, the OBC non-Bloch bulk invariant $\nu_{\rm OBC}$ jumps from $1$ to $0$ [Fig.~\ref{fig3}(f)], marking the disappearance of the zero-energy boundary modes. At the same point, the average skin exponent $\bar{\kappa}$, defined as the mean of the skin exponents evaluated along the selected GBZ contour, changes sign [Fig.~\ref{fig3}(g)], indicating a reversal of the dominant skin-accumulation direction. Simultaneously, the PBC point-gap winding number $\nu_{\rm PBC}$ switches from $-1$ to $+1$ [Fig.~\ref{fig3}(h)]. These concurrent changes provide direct numerical evidence that, beyond the solvable limit, an EP-constrained path can lock the OBC non-Bloch bulk real-line-gap transition to the monitored PBC point-gap winding transition within the present SSH-type model. Representative finite-size checks for the two-band model are provided in Appendix~\ref{app:finite_size}.

\subsection{Breakdown of locking for unconstrained paths}

\begin{figure}[htbp]
	\centering
	\includegraphics[width=\columnwidth]{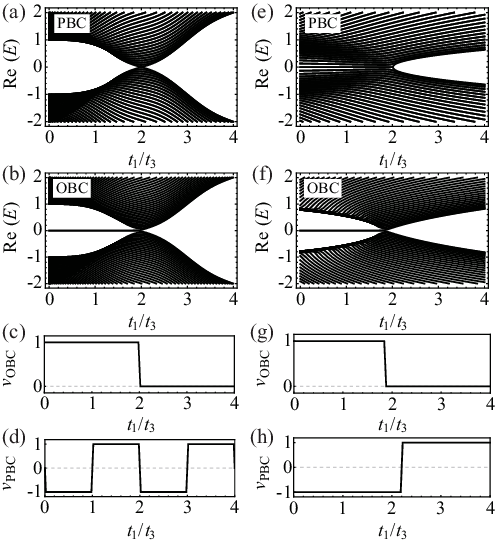}
	\caption{Breakdown of locking outside EP-constrained paths for $\delta=0.5t_3$. 
	(a)--(d) Results along the sweep $t_2=t_1+0.25\,t_3\sin[(t_1/t_3-1)\pi]$: 
	(a) real parts of the PBC spectrum, 
	(b) real parts of the finite-chain OBC spectrum, 
	(c) OBC non-Bloch bulk real-line-gap invariant $\nu_{\rm OBC}$, and 
	(d) PBC point-gap winding number $\nu_{\rm PBC}$. 
	(e)--(h) Results for the sweep with fixed $t_2=2.2t_3$: 
	(e) real parts of the PBC spectrum, 
	(f) real parts of the finite-chain OBC spectrum, 
	(g) $\nu_{\rm OBC}$, and 
	(h) $\nu_{\rm PBC}$. 
	All quantities are plotted as functions of $t_1/t_3$.}
	\label{fig4}
\end{figure}

We now show that the locking is lost when the parameter sweep is not confined to an EP-constrained manifold. This distinction is essential: neither an isolated exceptional point nor an isolated Hermitian degeneracy is sufficient to synchronize the PBC point-gap winding transition with the OBC non-Bloch bulk real-line-gap transition.

Figure~\ref{fig4} illustrates this breakdown for two representative unconstrained sweeps at $\delta=0.5t_3$. We first consider the path
\begin{equation}
	t_2=t_1+0.25t_3\sin\bigl[(t_1/t_3-1)\pi\bigr],
\end{equation}
shown in Figs.~\ref{fig4}(a)--\ref{fig4}(d). Along this sweep, the Bloch spectrum is not constrained to remain pinned to a zero-energy exceptional degeneracy. As shown in Fig.~\ref{fig4}(a), the PBC spectrum does not exhibit a persistent zero-energy EP throughout the evolution. At $t_1=2t_3$, one has $t_2=t_1$, so the system passes through the Hermitian limit, where the zero-energy touching is an ordinary nondefective degeneracy rather than an EP. Under OBCs, the corresponding non-Bloch bulk real line gap closes only once, at $t_1=2t_3$. This is reflected in the finite-chain OBC spectrum in Fig.~\ref{fig4}(b) and in the jump of the non-Bloch invariant in Fig.~\ref{fig4}(c), where $\nu_{\rm OBC}$ changes from $1$ for $t_1<2t_3$ to $0$ for $t_1>2t_3$. By contrast, the PBC point-gap winding number changes repeatedly, at $t_1/t_3=1$, $2$, and $3$, as shown in Fig.~\ref{fig4}(d). Thus, the isolated Hermitian degeneracy at $t_1=2t_3$ does not enforce locking: it produces at most an accidental coincidence between the OBC non-Bloch bulk real-line-gap transition and one of several PBC point-gap winding transitions.

We next consider the second unconstrained path, with fixed $t_2=2.2t_3$, shown in Figs.~\ref{fig4}(e)--\ref{fig4}(h). In this case, the PBC spectrum encounters a zero-energy EP only at an isolated parameter value, so the sweep is not confined to an EP-constrained manifold. Figures~\ref{fig4}(f) and \ref{fig4}(g) show that the OBC non-Bloch bulk real line gap closes at $t_1=1.8t_3$, where the non-Bloch invariant changes. However, Fig.~\ref{fig4}(h) shows that the PBC point-gap winding number changes instead at $t_1=2.2t_3$. The two critical points are therefore distinct. This example demonstrates that the mere presence of isolated zero-energy EPs along a parameter path is not sufficient for locking. What locks the transitions in the EP-constrained sweeps is the persistence of the zero-energy Bloch-degeneracy constraint along the entire parameter evolution.

Taken together, the constrained and unconstrained sweeps establish the role of EP-constrained evolution within the present SSH-type model. The locking between the OBC non-Bloch bulk real-line-gap transition and the PBC point-gap winding transition is not generic, nor is it implied by isolated EPs or isolated Hermitian degeneracies. Rather, it requires the entire parameter sweep to remain on an EP-constrained manifold, so that the zero-energy Bloch degeneracy is maintained continuously along the path. We next show that the same principle persists in a four-band extension.

\section{Multiband extension with branch-resolved GBZs}

We next examine whether the EP-constrained locking mechanism persists in a multiband setting. This question is nontrivial because, under open boundary conditions, different spectral branches can undergo distinct non-Bloch deformations and therefore be associated with different GBZs. In such cases, the OBC non-Bloch bulk topology must be evaluated using branch-resolved non-Bloch loops rather than a single common GBZ.

To test this issue, we consider a spinful four-band extension of the non-Hermitian SSH chain. Although the spectral branches originate from the same microscopic lattice, they generally exhibit inequivalent non-Bloch root structures. Consequently, the corresponding GBZs and skin exponents can differ substantially from one branch to another. This provides a useful setting for testing whether the EP-constrained locking mechanism identified in the two-band model remains operative when the open-boundary bulk spectrum is described by branch-resolved GBZs.

\begin{figure*}[htbp]
  \centering
  \includegraphics[width=0.9\textwidth]{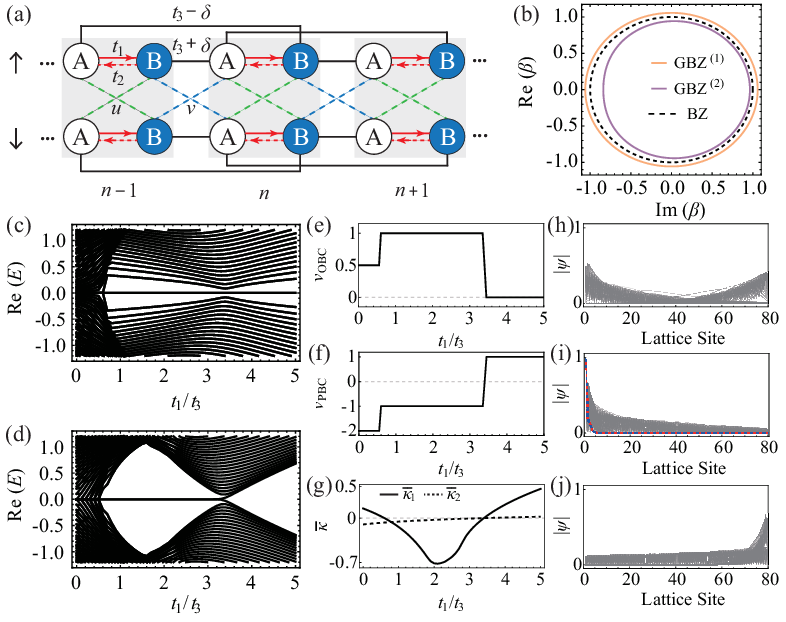}
  \caption{Four-band spinful extension of the non-Hermitian SSH model. 
  (a) Schematic of the spinful non-Hermitian SSH ladder, showing the intracell and intercell hoppings together with the intracell and intercell spin-flip couplings $u$ and $v$. 
  (b) Branch-resolved GBZ loops for the two chiral branches at $t_1=0.4t_3$. 
  (c) Real parts of the PBC spectrum and 
  (d) real parts of the finite-chain OBC spectrum as functions of $t_1/t_3$. 
  (e) OBC non-Bloch bulk real-line-gap invariant $\nu_{\rm OBC}$. 
  (f) PBC point-gap winding number $\nu_{\rm PBC}$. 
  (g) Branch-resolved average skin exponents $\bar{\kappa}_1$ and
$\bar{\kappa}_2$, defined in Eq.~\eqref{eq:branch_avg_kappa}. 
  (h)--(j) OBC eigenstate profiles at $t_1=0.4t_3$, $1.7t_3$, and $3.6t_3$, respectively. Bulk skin modes are shown in gray, while zero-energy boundary modes, when present, are highlighted in blue and red. 
  Unless otherwise specified, the parameters are $t_2=3.4t_3$, $\delta=-0.5t_3$, $u=2t_3$, and $v=0.6t_3$.}
\label{fig5}
\end{figure*}

\begin{figure}[htbp]
  \centering
  \includegraphics[width=\columnwidth]{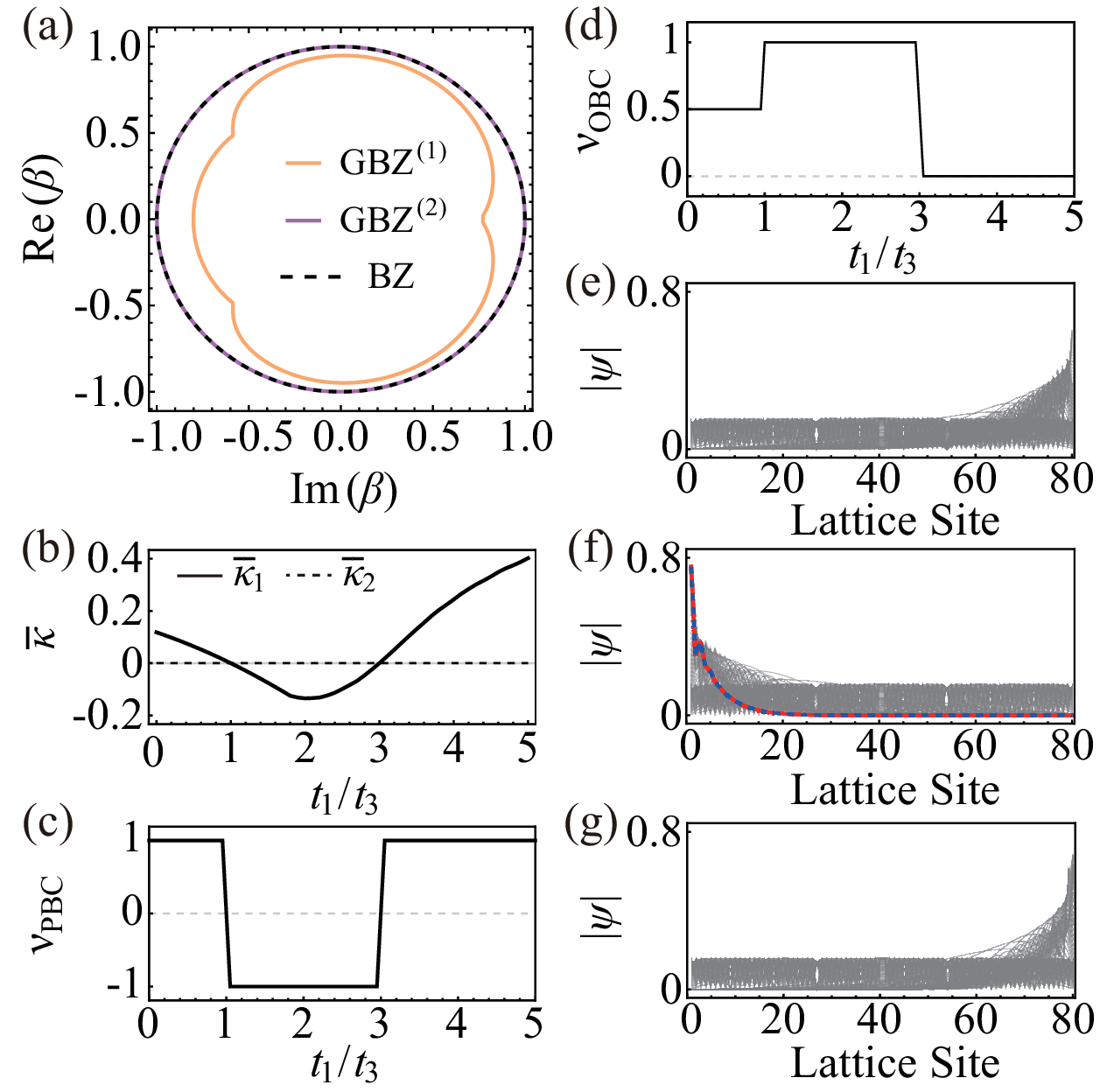}
  \caption{Locked transitions along an EP-constrained sweep in a branch-imbalanced regime. 
  (a) Branch-resolved GBZ loops at $t_1=2.6t_3$, with one branch strongly deformed away from the unit circle and the other remaining close to it. 
  (b) Branch-resolved average skin exponents $\bar{\kappa}_1$ and
$\bar{\kappa}_2$, defined in Eq.~\eqref{eq:branch_avg_kappa}.
  (c), (d) PBC point-gap winding number $\nu_{\rm PBC}$ and OBC non-Bloch bulk real-line-gap invariant $\nu_{\rm OBC}$ as functions of $t_1/t_3$, exhibiting locked critical points at $t_1/t_3=1$ and $3$. 
  (e)--(g) OBC eigenstate profiles at $t_1=0.4t_3$, $2.6t_3$, and $3.4t_3$, respectively. Bulk modes are shown in gray, while zero-energy boundary modes, when present, are highlighted in blue and red.
  The parameters are $t_2=3t_3$, $\delta=-0.5t_3$, $u=2t_3$, and $v=t_3$.}
\label{fig6}
\end{figure}

\subsection{Spinful four-band model and branch-resolved GBZs}

As shown schematically in Fig.~\ref{fig5}(a), we consider a spinful four-band extension of the non-Hermitian SSH chain. The model consists of two spin-resolved copies of the extended non-Hermitian SSH chain coupled by spin-flip hopping terms. The real-space Hamiltonian for this four-band model is written as
\begin{equation}
	H=H_{0}+H_{\mathrm{s}},
\end{equation}
where
\begin{align}
  H_{0}=&\sum_{n,\sigma}\Big[t_1\, b_{n,\sigma}^{\dagger}a_{n,\sigma}
    +t_2\, a_{n,\sigma}^{\dagger}b_{n,\sigma}\notag \\
   & +(t_3+\delta)\bigl(b_{n,\sigma}^{\dagger}a_{n+1,\sigma}+\mathrm{h.c.}\bigr)\notag \\
   & +(t_3-\delta)\bigl(a_{n,\sigma}^{\dagger}b_{n+1,\sigma}+\mathrm{h.c.}\bigr)\Big],
  \label{eq:four_band_h0}
\end{align}
and
\begin{equation}
H_{\mathrm{s}}=\sum_{n,\sigma}\Big[u\bigl(a_{n,\sigma}^{\dagger}b_{n,\bar{\sigma}}+\mathrm{h.c.}\bigr)
+v\bigl(a_{n+1,\sigma}^{\dagger}b_{n,\bar{\sigma}}+\mathrm{h.c.}\bigr)\Big].
  \label{eq:spin_flip_real_space}
\end{equation}
Here $\sigma=\uparrow,\downarrow$, $\bar{\sigma}$ denotes the opposite spin, and $u$ and $v$ are the intracell and intercell spin-flip hopping amplitudes, respectively. Since all hopping terms connect opposite sublattices, the four-band model preserves the chiral structure of the two-band problem. In momentum space, the spin-flip hopping amplitude is denoted by $h_{\rm s}(k)=u+ve^{ik}$; the explicit block form of the Bloch Hamiltonian is given in Appendix~\ref{app:four_band}.

Under OBCs, the two chiral spectral branches generally acquire distinct non-Bloch deformations. They are therefore described by branch-resolved GBZ loops, denoted by ${\rm GBZ}^{(1)}$ and ${\rm GBZ}^{(2)}$, rather than by a single common GBZ. As a result, different branches can contribute unequally to the NHSE, and strongly branch-imbalanced skin localization may occur. The corresponding open-boundary bulk topology must then be evaluated from the branch-resolved non-Bloch spectra.

The explicit Bloch and non-Bloch Hamiltonians, the construction of the branch-resolved GBZs, and the definition of the OBC non-Bloch bulk real-line-gap invariant are given in Appendix~\ref{app:four_band}. Here we focus on the resulting spectral evolution and test whether EP-constrained sweeps continue to lock the PBC point-gap winding transitions to the OBC non-Bloch bulk real-line-gap transitions in this multiband setting. Unless otherwise stated, the PBC winding number in each four-band sweep is evaluated with respect to a fixed reference energy chosen inside the monitored point-gap region and kept unchanged throughout that sweep.

\subsection{EP-constrained locking with branch-dependent skin effects}

We first consider an EP-constrained sweep in a generic four-band regime, where both chiral branches exhibit appreciable non-Bloch deformation. A representative example is shown in Fig.~\ref{fig5} for $t_2=3.4t_3$, $\delta=-0.5t_3$, $u=2t_3$, and $v=0.6t_3$, with $t_1/t_3$ varied along the sweep. For this parameter choice, one chiral branch remains pinned to a zero-energy degeneracy throughout the evolution. Indeed, at $k=\pi$ one has
\begin{equation}
	h_2(\pi)-h_{\rm s}(-\pi)	=	(t_2-2t_3)-(u-v)=0,
  \label{eq:fourbandEP}
\end{equation}
so that the first chiral branch, as defined in Eq.~\eqref{eq:four_band_branches}, contains a zero-energy Bloch degeneracy for all values of $t_1/t_3$. The parameter path therefore stays on an EP-constrained manifold.

The branch-resolved GBZs at $t_1=0.4t_3$ are shown in Fig.~\ref{fig5}(b). The loop ${\rm GBZ}^{(1)}$ lies predominantly outside the unit circle, whereas ${\rm GBZ}^{(2)}$ lies predominantly inside it. This indicates opposite skin-localization tendencies for the two chiral branches. Despite this strong branch dependence, the locking mechanism remains intact.

Figures~\ref{fig5}(c) and \ref{fig5}(d) show the real parts of the PBC spectrum and the finite-chain OBC spectrum, respectively, as $t_1/t_3$ is varied. Under PBCs, the EP constraint is manifested by a zero-energy Bloch degeneracy that
remains pinned at $\mathrm{Re}(E)=0$ throughout the sweep, as shown in Fig.~\ref{fig5}(c). The associated OBC response is captured by the branch-resolved non-Bloch bulk continuum. For $0\le t_1/t_3<0.6$, the non-Bloch bulk spectrum is real-line-gapless because the bulk continuum intersects the reference line $\mathrm{Re}(E)=0$. At $t_1/t_3=0.6$, the continuum detaches from
this reference line, a non-Bloch bulk real line gap opens, and zero-energy boundary modes appear in the finite open chain, marking the onset of a topological line-gapped phase. Upon further increasing $t_1$, the continuum touches $\mathrm{Re}(E)=0$ again at $t_1/t_3=3.4$, where the real line gap closes
and then reopens into a topologically trivial line-gapped phase without zero-energy boundary modes.

This evolution is captured by the OBC non-Bloch bulk real-line-gap invariant $\nu_{\rm OBC}$ in Fig.~\ref{fig5}(e). In the interval $0\le t_1/t_3<0.6$, where the non-Bloch bulk real line gap is absent, the nonquantized value of $\nu_{\rm OBC}$ should be regarded only as a numerical continuation and carries no topological meaning~\cite{PhysRevLett.121.086803,Xiao2020}. Once the real line gap opens, $\nu_{\rm OBC}$ becomes quantized at $1$, identifying a topological line-gapped phase with zero-energy boundary modes. At $t_1/t_3=3.4$, $\nu_{\rm OBC}$ drops from $1$ to $0$, marking the transition to a trivial line-gapped phase. Crucially, the PBC point-gap winding number changes at the same critical values, as shown in Fig.~\ref{fig5}(f). It changes from $-2$ to $-1$ at $t_1/t_3=0.6$, where the OBC non-Bloch bulk real line gap opens, and from $-1$ to $+1$ at $t_1/t_3=3.4$, where the OBC non-Bloch bulk real line gap closes and reopens with different topology. Thus, both the entrance into the topological line-gapped phase and the exit from it are synchronized with changes in the PBC point-gap winding.

To characterize the branch-dependent skin effect, we introduce the branch-resolved average skin exponent
\begin{equation}
	\bar{\kappa}_{\eta}=
	\frac{1}{2\pi}\int_0^{2\pi} d\theta\,
	\ln|\beta_{\eta}(\theta)|,
	\label{eq:branch_avg_kappa}
\end{equation}
where $\eta=1,2$ labels the two GBZ branches and $\beta_{\eta}(\theta)$ denotes the corresponding branch-resolved GBZ trajectory. In the present multiband setting, $|\nu_{\rm PBC}|=2$ is associated with bipolar skin accumulation, whereas $|\nu_{\rm PBC}|=1$ corresponds to predominantly unipolar skin accumulation in the representative spectra studied here. This interpretation is supported by the
branch-resolved average skin exponents $\bar{\kappa}_1$ and $\bar{\kappa}_2$ in Fig.~\ref{fig5}(g), whose signs and magnitudes change across the same critical values. The corresponding real-space eigenstate profiles in Figs.~\ref{fig5}(h)--\ref{fig5}(j) are fully consistent with this picture: the bulk states exhibit branch-dependent skin accumulation, while the zero-energy boundary modes appear only in the topological line-gapped regime.

These results show that the EP-constrained locking mechanism persists in this four-band extension, even though the open-boundary bulk spectrum must be described by two branch-resolved GBZs with distinct skin exponents. Thus, within this multiband SSH-type model, an EP-constrained sweep continues to lock the OBC non-Bloch bulk real-line-gap transitions to the corresponding changes in the PBC point-gap winding.

\subsection{EP-constrained locking with branch-imbalanced non-Bloch response}

We next consider a more stringent situation in which the non-Bloch response is strongly branch imbalanced: one spectral branch is substantially deformed, while the other remains close to the conventional Brillouin zone. A representative example is shown in Fig.~\ref{fig6} for $t_2=3t_3$, $\delta=-0.5t_3$, $u=2t_3$, and $v=t_3$ ($v=-2\delta$), with $t_1/t_3$ varied along the sweep. As in the previous example, the condition Eq.~\eqref{eq:fourbandEP} is satisfied. Hence one chiral branch remains pinned to a zero-energy Bloch degeneracy throughout the evolution, and the parameter path stays on an EP-constrained manifold.

The branch-resolved non-Bloch geometry is shown in Fig.~\ref{fig6}(a). One GBZ branch ${\rm GBZ}^{(1)}$ is strongly deformed away from the unit circle, whereas the other ${\rm GBZ}^{(2)}$ stays close to the conventional Brillouin zone (BZ). This demonstrates a pronounced asymmetry in the branch-resolved non-Bloch response. The imbalance is quantified in Fig.~\ref{fig6}(b), where the average skin exponent $\bar{\kappa}_1$ gives the dominant contribution, while $\bar{\kappa}_2$ remains close to zero over most of the sweep.

Despite this strong branch imbalance, the topological transitions remain locked. As shown in Figs.~\ref{fig6}(c) and \ref{fig6}(d), the PBC point-gap winding number $\nu_{\rm PBC}$ and the OBC non-Bloch bulk real-line-gap invariant $\nu_{\rm OBC}$ change at the same critical values, $t_1/t_3=1$ and $3$. Thus, the opening and closing of the OBC non-Bloch bulk real line gap are synchronized with the corresponding changes in the PBC point-gap winding, even though the non-Bloch deformation is carried predominantly by only one branch.

The corresponding real-space eigenstate profiles in Figs.~\ref{fig6}(e)--\ref{fig6}(g) confirm this picture. In the line-gapless regime, the bulk states exhibit skin accumulation, but no zero-energy boundary modes are present. In the intermediate topological line-gapped regime, zero-energy boundary modes appear, accompanied by a reversal of the dominant skin-localization tendency. Beyond the second transition, the zero modes disappear, and the bulk states again accumulate predominantly toward the opposite boundary. Therefore, even when the non-Bloch response is concentrated mainly in a single branch, an EP-constrained sweep still synchronizes the OBC non-Bloch bulk real-line-gap transition with the corresponding change in the PBC point-gap winding.

The results in Figs.~\ref{fig5} and \ref{fig6} demonstrate that the EP-constrained locking mechanism is not restricted to the minimal two-band setting. Within the present four-band SSH-type extension, it persists in the presence of branch-resolved GBZs and remains effective even in strongly branch-imbalanced regimes, where the open-boundary non-Bloch response is highly asymmetric across spectral branches.

\section{Conclusion}
\label{sec:conclusion}

In conclusion, we have identified EP-constrained evolution as a path-dependent mechanism for locking PBC point-gap winding transitions to OBC non-Bloch bulk real-line-gap transitions in the chiral SSH-type non-Hermitian lattices studied here. The key ingredient is not the mere presence of EPs, but the persistence of a zero-energy Bloch-degeneracy constraint along the entire parameter sweep. Under this constraint, changes in the PBC point-gap winding coincide with OBC non-Bloch bulk real-line-gap transitions and with the associated appearance or disappearance of zero-energy boundary modes.

This result clarifies that chiral symmetry and isolated degeneracies alone do not enforce such locking. Chiral symmetry pins the relevant degeneracies to zero energy, but unconstrained sweeps show that isolated EPs or Hermitian band touchings are insufficient to synchronize PBC and OBC criticalities. Thus, within the SSH-type class studied here, EP-constrained paths provide a path-dependent diagnostic principle for relating monitored PBC point-gap winding changes to OBC non-Bloch bulk real-line-gap transitions, without implying a generic equivalence between point-gap and line-gap topology.

We established this mechanism analytically in a solvable limit of the extended non-Hermitian SSH model and numerically beyond this limit using GBZ calculations. We further showed that the locking persists in a spinful four-band extension with branch-resolved GBZs, including regimes with strongly branch-imbalanced skin responses. These results suggest that PBC spectral winding can diagnose OBC non-Bloch bulk topological transitions along appropriately constrained parameter paths. Future directions include extending this principle to higher-dimensional non-Hermitian band structures~\cite{PhysRevLett.121.136802}, Floquet systems~\cite{park2022revealing}, other symmetry classes, and settings with disorder, interactions, finite-size effects, or measurement-induced phenomena~\cite{li2021impurity,PhysRevB.108.184205,PhysRevB.102.035153,li2025size,qin2026many,liu2025entanglement,PhysRevLett.133.136602}. It would also be useful to explore implementations in platforms where complex spectral winding or non-Bloch topology can be probed directly~\cite{PhysRevLett.124.250402,zhao2025two,wang2025topological,wang2021generating,PhysRevLett.130.163001}.

\begin{acknowledgments}
	Z. X. is supported by Quantum Science and Technology-National Science and Technology Major Project (Grant No. 2025ZD0300400), the NSFC (Grant Nos. 12375016 and 12461160324), and Beijing National Laboratory for Condensed Matter Physics (No. 2023BNLCMPKF001).
\end{acknowledgments}

\section*{Data Availability}
The data that support the findings of this article are available from the authors upon reasonable request.

\appendix

\section{Non-Bloch formulation and GBZ construction for the two-band model}
\label{app:two_band}

Under OBCs, translational invariance is lost, and bulk eigenstates generally acquire exponential spatial envelopes. In the thermodynamic limit, such generalized Bloch states can be described by replacing $e^{ik}$ with a complex variable $\beta$, leading to the non-Bloch Hamiltonian
\begin{equation}
	H(\beta)=
	\begin{bmatrix}
		0 & h_{2}(\beta) \\
		h_{1}(\beta) & 0
	\end{bmatrix},
	\label{eq:non_bloch_hamiltonian}
\end{equation}
where
\begin{align}
	h_1(\beta)&=t_1+(t_3+\delta)\beta+(t_3-\delta)\beta^{-1},\notag\\
	h_2(\beta)&=t_2+(t_3+\delta)\beta^{-1}+(t_3-\delta)\beta.
	\label{eq:non_bloch_offdiagonal}
\end{align}
The non-Bloch spectrum is determined by the characteristic equation
\begin{equation}
	h_1(\beta)h_2(\beta)=E^2.
	\label{eq:two_band_characteristic}
\end{equation}
For a given complex energy $E$, this equation generically gives four roots after clearing the negative powers of $\beta$. We order them as
\begin{equation}
	|\beta_1(E)|\le |\beta_2(E)|\le |\beta_3(E)|\le |\beta_4(E)|.
\end{equation}
The GBZ is then obtained from the equal-modulus condition~\cite{PhysRevLett.123.066404,10.1093/ptep/ptaa140}
\begin{equation}
	|\beta_2(E)|=|\beta_3(E)|.
	\label{eq:gbz_equal_modulus}
\end{equation}
The OBC non-Bloch bulk spectrum is generated by evaluating $H(\beta)$ on this GBZ, and the non-Bloch invariant $\nu_{\rm OBC}$ is computed along the same contour when the corresponding real line gap is open. In the analytically tractable limit $\delta=t_3$, the GBZ reduces to a circle with radius
\begin{equation}
	|\beta|=\sqrt{\left|t_1/t_2\right|}.
\end{equation}

\section{Four-band model and branch-resolved GBZ construction}
\label{app:four_band}

From the real-space Hamiltonians in Eqs.~\eqref{eq:four_band_h0} and \eqref{eq:spin_flip_real_space}, the Bloch Hamiltonian in the basis $(a_\uparrow,b_\uparrow,a_\downarrow,b_\downarrow)^T$ can be written as
\begin{equation}
	H(k)=s_{0}\otimes H_{0}(k)+s_{x}\otimes H_{\mathrm{s}}(k),
	\label{eq:four_band_bloch}
\end{equation}
where $s_0$ is the identity matrix in the spin space and $s_x$ is the corresponding Pauli matrix; $\tau_\mu$ will denote Pauli matrices acting in the sublattice space. The two sublattice-space blocks are
\begin{equation}
	H_{0}(k)=
	\begin{bmatrix}
		0 & h_{2}(k) \\
		h_{1}(k) & 0
	\end{bmatrix},\qquad
	H_{\mathrm{s}}(k)=
	\begin{bmatrix}
		0 & h_{\mathrm{s}}(-k) \\
		h_{\mathrm{s}}(k) & 0
	\end{bmatrix},
	\label{eq:four_band_blocks}
\end{equation}
with $h_{1}(k)$ and $h_{2}(k)$ defined in Eq.~\eqref{eq:bloch_offdiagonal}, and
\begin{equation}
	h_{\mathrm{s}}(k)=u+ve^{ik}.
\end{equation}
The Hamiltonian preserves chiral symmetry,
\begin{equation}
	\Gamma H(k)\Gamma^{-1}=-H(k),\qquad
	\Gamma=s_0\otimes \tau_z,
	\label{eq:four_band_chiral_symmetry}
\end{equation}
where $\tau_z$ acts in the sublattice space.

Since the spin-flip term is proportional to $s_x$, the Hamiltonian can be block diagonalized in the eigenbasis of $s_x$. It therefore separates into two effective chiral branches,
\begin{align}
	E_{\pm}^{(1)}(k)
	&=\pm\sqrt{\bigl[h_{1}(k)-h_{\mathrm{s}}(k)\bigr]
		\bigl[h_{2}(k)-h_{\mathrm{s}}(-k)\bigr]},\notag \\
	E_{\pm}^{(2)}(k)
	&=\pm\sqrt{\bigl[h_{1}(k)+h_{\mathrm{s}}(k)\bigr]
		\bigl[h_{2}(k)+h_{\mathrm{s}}(-k)\bigr]}.
	\label{eq:four_band_branches}
\end{align}
Here the branch labels correspond to the two eigenvalues of $s_x$, with the sign convention chosen as in Eq.~\eqref{eq:four_band_branches}.

Under OBCs, the non-Bloch formulation is obtained by replacing $e^{ik}$ with a complex parameter $\beta$. The resulting non-Bloch Hamiltonian reads
\begin{equation}
	H(\beta)=s_0\otimes H_0(\beta)+s_x\otimes H_{\mathrm{s}}(\beta),
	\label{eq:four_band_non_bloch}
\end{equation}
where $h_{1}(\beta)$ and $h_{2}(\beta)$ are given in Eq.~\eqref{eq:non_bloch_offdiagonal}, and
\begin{equation}
	H_{\mathrm{s}}(\beta)=
	\begin{bmatrix}
		0 & h_{\mathrm{s}}(\beta^{-1}) \\
		h_{\mathrm{s}}(\beta) & 0
	\end{bmatrix},\qquad
	h_{\mathrm{s}}(\beta)=u+v\beta .
\end{equation}
For a fixed complex energy $E$ on branch $\eta=1,2$, the branch-resolved characteristic equations are
\begin{align}
	E^{2}
	&=\bigl[h_{1}(\beta)-h_{\mathrm{s}}(\beta)\bigr]
	\bigl[h_{2}(\beta)-h_{\mathrm{s}}(\beta^{-1})\bigr],
	\qquad \eta=1,\notag \\
	E^{2}
	&=\bigl[h_{1}(\beta)+h_{\mathrm{s}}(\beta)\bigr]
	\bigl[h_{2}(\beta)+h_{\mathrm{s}}(\beta^{-1})\bigr],
	\qquad \eta=2.
	\label{eq:four_band_branch_equations}
\end{align}
After clearing the negative powers of $\beta$, each equation generically gives four roots. For branch $\eta$, we order them as
\begin{equation}
	|\beta_1^{(\eta)}(E)|\le
	|\beta_2^{(\eta)}(E)|\le
	|\beta_3^{(\eta)}(E)|\le
	|\beta_4^{(\eta)}(E)|.
\end{equation}
The branch-resolved GBZ is then determined by the equal-modulus condition
\begin{equation}
	|\beta_2^{(\eta)}(E)|=|\beta_3^{(\eta)}(E)|,
	\label{eq:four_band_gbz_condition}
\end{equation}
which defines the contour ${\rm GBZ}^{(\eta)}$. Thus, the four-band model supports two branch-resolved GBZ loops, denoted by ${\rm GBZ}^{(1)}$ and ${\rm GBZ}^{(2)}$.

Because the four-band Hamiltonian decomposes into two effective chiral branches in the $s_x$ eigenbasis, the OBC invariant can be evaluated branch by branch. When the OBC non-Bloch bulk real line gap is open, the total multiband invariant is obtained by summing the branch-resolved biorthogonal Berry phases,
\begin{equation}
	\nu_{\rm OBC}=\sum_{\eta=1,2}\frac{\varphi_B^{(\eta)}}{2\pi},
	\qquad
	\varphi_B^{(\eta)}=\varphi_{Z+}^{(\eta)}+\varphi_{Z-}^{(\eta)}.
	\label{eq:four_band_obc_invariant}
\end{equation}
Here
\begin{equation}
	\varphi_{Z\pm}^{(\eta)}
	=-\oint_{{\rm GBZ}^{(\eta)}} d\theta\,
	\frac{
	\langle \psi_{\pm}^{(\eta,L)}(\theta)|\, i\,\partial_{\theta}\,|\psi_{\pm}^{(\eta,R)}(\theta)\rangle
	}{
	\langle \psi_{\pm}^{(\eta,L)}(\theta)|\psi_{\pm}^{(\eta,R)}(\theta)\rangle
	},
	\label{eq:four_band_berry_phase}
\end{equation}
where $|\psi_{\pm}^{(\eta,R)}(\theta)\rangle$ and $|\psi_{\pm}^{(\eta,L)}(\theta)\rangle$ are the right and left non-Bloch bulk eigenstates evaluated along ${\rm GBZ}^{(\eta)}$. The invariant is quantized only when the OBC non-Bloch bulk real line gap is open.

\section{Finite-size stability of OBC zero modes in the two- and four-band models}
\label{app:finite_size}

We further examine the finite-size stability of the zero-energy boundary modes in representative two-band and four-band parameter sweeps. For each finite open chain, we compute the minimum absolute OBC eigenenergy,
\begin{equation}
	\epsilon_{\min}(N)=\min_n |E_n^{\rm OBC}|,
\end{equation}
where $E_n^{\rm OBC}$ denotes the $n$-th OBC eigenvalue for an open chain with $N$ unit cells. A near-zero value of $\epsilon_{\min}(N)$ provides a finite-size indicator of zero-energy boundary modes, whereas a value remaining finite with increasing $N$ indicates their absence.

\begin{figure}[htbp]
	\centering
	\includegraphics[width=0.8\columnwidth]{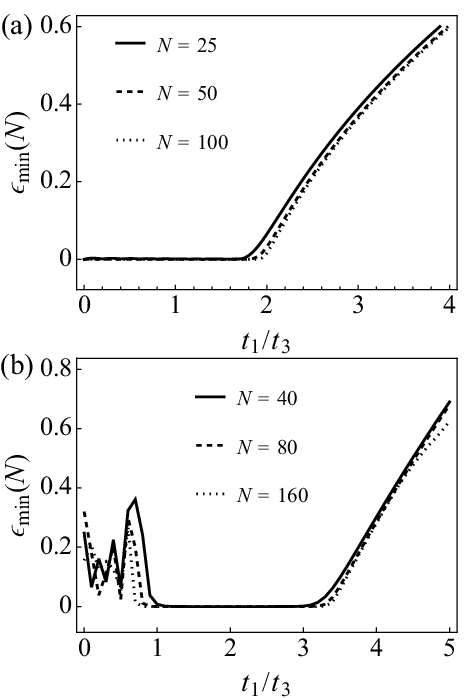}
	\caption{Finite-size stability of OBC zero modes in the two-band and four-band models. The minimum absolute OBC eigenenergy $\epsilon_{\min}(N)$ is plotted as a function of $t_1/t_3$. (a) Two-band model with $t_2=2t_3$ and $\delta=0.5t_3$ for $N=25$, $50$, and $100$. (b) Four-band model with $t_2=3.4t_3$, $\delta=-0.5t_3$, $u=2t_3$, and $v=0.6t_3$ for $N=40$, $80$, and $160$. Near-zero values of $\epsilon_{\min}(N)$ indicate the presence of zero-energy boundary modes.}
	\label{fig7}
\end{figure}

Figure~\ref{fig7}(a) shows the result for the two-band model with $t_2=2t_3$ and $\delta=0.5t_3$, corresponding to the representative sweep in Fig.~\ref{fig3}. For $N=25$, $50$, and $100$, the zero-mode regime and the transition near $t_1/t_3=2$ remain stable as the system size increases, with only finite-size rounding close to the transition.

Figure~\ref{fig7}(b) shows the corresponding result for the four-band model with $t_2=3.4t_3$, $\delta=-0.5t_3$, $u=2t_3$, and $v=0.6t_3$, corresponding to the sweep in Fig.~\ref{fig5}. For $N=40$, $80$, and $160$, the intermediate zero-mode regime is likewise stable against increasing system size. The finite-size data also reproduce the expected transition boundaries of the topological line-gapped regime. These results support the robustness of the OBC zero-mode regimes in both the two-band and four-band models.

\section{Numerical procedures for spectra, GBZs, and topological invariants}
\label{app:numerics}

Unless otherwise specified in the figure captions, OBC spectra and real-space eigenstate profiles were obtained by exact diagonalization of finite chains with $N=50$ unit cells for the two-band model and $N=80$ unit cells for the four-band model. PBC spectra were evaluated on uniform momentum meshes. The winding numbers and the locations of the critical points were checked for convergence by increasing the mesh density. For each parameter sweep, the reference energy $E_{\mathrm{b}}$ in Eq.~\eqref{eq:pbc_winding} was chosen inside the relevant point-gap region and then kept fixed throughout the sweep. The same value of $E_{\mathrm{b}}$ was used for all data points within a given sweep. In the figures, $E_{\mathrm{b}}$ was chosen away from the pinned zero-energy Bloch degeneracy so that the monitored point gap remained well defined except at the indicated PBC winding transitions.

Except in the analytically tractable limit $\delta=t_3$, where the GBZ is known in closed form, the GBZs were constructed numerically by solving the characteristic equation for $\beta$ and continuously tracking the roots satisfying the non-Bloch equal-modulus condition. In the four-band model, this procedure was performed separately for the two chiral branches, yielding the branch-resolved GBZ loops. The biorthogonal Berry phases were evaluated on discretized GBZ contours using left and right non-Bloch eigenvectors normalized with respect to the biorthogonal inner product. When the OBC non-Bloch bulk spectrum touched $\mathrm{Re}(E)=0$ and the real line gap closed, the resulting value of $\nu_{\rm OBC}$ was treated only as a numerical continuation, rather than as a well-defined topological invariant.

The skin exponent associated with a non-Bloch factor $\beta$ was defined as $\kappa=\ln|\beta|$. In the multiband calculations, branch-resolved averages were evaluated separately on the two chiral GBZ branches. The continuous average in Eq.~\eqref{eq:branch_avg_kappa} was implemented numerically as
\begin{equation}
	\bar{\kappa}_{\eta}	=	\frac{1}{N_{\rm GBZ}}\sum_{j=1}^{N_{\rm GBZ}}
	\ln|\beta_{\eta,j}|,
	\label{eq:average_skin_exponent}
\end{equation}
where $\eta=1,2$ labels the branch, $N_{\rm GBZ}$ is the number of discretization points on the selected GBZ loop, and $\beta_{\eta,j}$ is evaluated along ${\rm GBZ}^{(\eta)}$. For the two-band model, the branch label is omitted.

\makeatletter
\immediate\write\@auxout{\string\citation{apsrev42Control}}
\makeatother


\begin{thebibliography}{68}%
	\makeatletter
	\providecommand \@ifxundefined [1]{%
		\@ifx{#1\undefined}
	}%
	\providecommand \@ifnum [1]{%
		\ifnum #1\expandafter \@firstoftwo
		\else \expandafter \@secondoftwo
		\fi
	}%
	\providecommand \@ifx [1]{%
		\ifx #1\expandafter \@firstoftwo
		\else \expandafter \@secondoftwo
		\fi
	}%
	\providecommand \natexlab [1]{#1}%
	\providecommand \enquote  [1]{``#1''}%
	\providecommand \bibnamefont  [1]{#1}%
	\providecommand \bibfnamefont [1]{#1}%
	\providecommand \citenamefont [1]{#1}%
	\providecommand \href@noop [0]{\@secondoftwo}%
	\providecommand \href [0]{\begingroup \@sanitize@url \@href}%
	\providecommand \@href[1]{\@@startlink{#1}\@@href}%
	\providecommand \@@href[1]{\endgroup#1\@@endlink}%
	\providecommand \@sanitize@url [0]{\catcode `\\12\catcode `\$12\catcode
		`\&12\catcode `\#12\catcode `\^12\catcode `\_12\catcode `\%12\relax}%
	\providecommand \@@startlink[1]{}%
	\providecommand \@@endlink[0]{}%
	\providecommand \url  [0]{\begingroup\@sanitize@url \@url }%
	\providecommand \@url [1]{\endgroup\@href {#1}{\urlprefix }}%
	\providecommand \urlprefix  [0]{URL }%
	\providecommand \Eprint [0]{\href }%
	\providecommand \doibase [0]{https://doi.org/}%
	\providecommand \selectlanguage [0]{\@gobble}%
	\providecommand \bibinfo  [0]{\@secondoftwo}%
	\providecommand \bibfield  [0]{\@secondoftwo}%
	\providecommand \translation [1]{[#1]}%
	\providecommand \BibitemOpen [0]{}%
	\providecommand \bibitemStop [0]{}%
	\providecommand \bibitemNoStop [0]{.\EOS\space}%
	\providecommand \EOS [0]{\spacefactor3000\relax}%
	\providecommand \BibitemShut  [1]{\csname bibitem#1\endcsname}%
	\let\auto@bib@innerbib\@empty
	%</preamble>
	\bibitem [{\citenamefont {Bender}\ and\ \citenamefont
		{Boettcher}(1998)}]{PhysRevLett.80.5243}%
	\BibitemOpen
	\bibfield  {author} {\bibinfo {author} {\bibfnamefont {C.~M.}\ \bibnamefont
			{Bender}}\ and\ \bibinfo {author} {\bibfnamefont {S.}~\bibnamefont
			{Boettcher}},\ }\bibfield  {title} {\bibinfo {title} {{Real spectra in
				{n}on-{H}ermitian Hamiltonians having PT symmetry}},\ }\href
	{https://doi.org/10.1103/PhysRevLett.80.5243} {\bibfield  {journal} {\bibinfo
			{journal} {Phys. Rev. Lett.}\ }\textbf {\bibinfo {volume} {80}},\ \bibinfo
		{pages} {5243} (\bibinfo {year} {1998})}\BibitemShut {NoStop}%
	\bibitem [{\citenamefont {Shen}\ \emph {et~al.}(2018)\citenamefont {Shen},
		\citenamefont {Zhen},\ and\ \citenamefont {Fu}}]{PhysRevLett.120.146402}%
	\BibitemOpen
	\bibfield  {author} {\bibinfo {author} {\bibfnamefont {H.}~\bibnamefont
			{Shen}}, \bibinfo {author} {\bibfnamefont {B.}~\bibnamefont {Zhen}},\ and\
		\bibinfo {author} {\bibfnamefont {L.}~\bibnamefont {Fu}},\ }\bibfield
	{title} {\bibinfo {title} {{Topological band theory for {n}on-{H}ermitian
				Hamiltonians}},\ }\href {https://doi.org/10.1103/PhysRevLett.120.146402}
	{\bibfield  {journal} {\bibinfo  {journal} {Phys. Rev. Lett.}\ }\textbf
		{\bibinfo {volume} {120}},\ \bibinfo {pages} {146402} (\bibinfo {year}
		{2018})}\BibitemShut {NoStop}%
	\bibitem [{\citenamefont {Bergholtz}\ \emph {et~al.}(2021)\citenamefont
		{Bergholtz}, \citenamefont {Budich},\ and\ \citenamefont
		{Kunst}}]{RevModPhys.93.015005}%
	\BibitemOpen
	\bibfield  {author} {\bibinfo {author} {\bibfnamefont {E.~J.}\ \bibnamefont
			{Bergholtz}}, \bibinfo {author} {\bibfnamefont {J.~C.}\ \bibnamefont
			{Budich}},\ and\ \bibinfo {author} {\bibfnamefont {F.~K.}\ \bibnamefont
			{Kunst}},\ }\bibfield  {title} {\bibinfo {title} {{Exceptional topology of
				{n}on-{H}ermitian systems}},\ }\href
	{https://doi.org/10.1103/RevModPhys.93.015005} {\bibfield  {journal}
		{\bibinfo  {journal} {Rev. Mod. Phys.}\ }\textbf {\bibinfo {volume} {93}},\
		\bibinfo {pages} {015005} (\bibinfo {year} {2021})}\BibitemShut {NoStop}%
	\bibitem [{\citenamefont {Ashida}\ \emph {et~al.}(2020)\citenamefont {Ashida},
		\citenamefont {Gong},\ and\ \citenamefont {Ueda}}]{ashida2020non}%
	\BibitemOpen
	\bibfield  {author} {\bibinfo {author} {\bibfnamefont {Y.}~\bibnamefont
			{Ashida}}, \bibinfo {author} {\bibfnamefont {Z.}~\bibnamefont {Gong}},\ and\
		\bibinfo {author} {\bibfnamefont {M.}~\bibnamefont {Ueda}},\ }\bibfield
	{title} {\bibinfo {title} {{{N}on-{H}ermitian physics}},\ }\href
	{https://doi.org/10.1080/00018732.2021.1876991} {\bibfield  {journal}
		{\bibinfo  {journal} {Adv. Phys.}\ }\textbf {\bibinfo {volume} {69}},\
		\bibinfo {pages} {249} (\bibinfo {year} {2020})}\BibitemShut {NoStop}%
	\bibitem [{\citenamefont {Schreiber}\ \emph {et~al.}(2011)\citenamefont
		{Schreiber}, \citenamefont {Cassemiro}, \citenamefont {Poto{\v{c}}ek},
		\citenamefont {G{\'a}bris}, \citenamefont {Jex},\ and\ \citenamefont
		{Silberhorn}}]{PhysRevLett.106.180403}%
	\BibitemOpen
	\bibfield  {author} {\bibinfo {author} {\bibfnamefont {A.}~\bibnamefont
			{Schreiber}}, \bibinfo {author} {\bibfnamefont {K.~N.}\ \bibnamefont
			{Cassemiro}}, \bibinfo {author} {\bibfnamefont {V.}~\bibnamefont
			{Poto{\v{c}}ek}}, \bibinfo {author} {\bibfnamefont {A.}~\bibnamefont
			{G{\'a}bris}}, \bibinfo {author} {\bibfnamefont {I.}~\bibnamefont {Jex}},\
		and\ \bibinfo {author} {\bibfnamefont {C.}~\bibnamefont {Silberhorn}},\
	}\bibfield  {title} {\bibinfo {title} {{Decoherence and disorder in quantum
				walks: From ballistic spread to localization}},\ }\href
	{https://doi.org/10.1103/PhysRevLett.106.180403} {\bibfield  {journal}
		{\bibinfo  {journal} {Phys. Rev. Lett.}\ }\textbf {\bibinfo {volume} {106}},\
		\bibinfo {pages} {180403} (\bibinfo {year} {2011})}\BibitemShut {NoStop}%
	\bibitem [{\citenamefont {Liang}\ \emph {et~al.}(2022)\citenamefont {Liang},
		\citenamefont {Xie}, \citenamefont {Dong}, \citenamefont {Li}, \citenamefont
		{Li}, \citenamefont {Gadway}, \citenamefont {Yi},\ and\ \citenamefont
		{Yan}}]{PhysRevLett.129.070401}%
	\BibitemOpen
	\bibfield  {author} {\bibinfo {author} {\bibfnamefont {Q.}~\bibnamefont
			{Liang}}, \bibinfo {author} {\bibfnamefont {D.}~\bibnamefont {Xie}}, \bibinfo
		{author} {\bibfnamefont {Z.}~\bibnamefont {Dong}}, \bibinfo {author}
		{\bibfnamefont {H.}~\bibnamefont {Li}}, \bibinfo {author} {\bibfnamefont
			{H.}~\bibnamefont {Li}}, \bibinfo {author} {\bibfnamefont {B.}~\bibnamefont
			{Gadway}}, \bibinfo {author} {\bibfnamefont {W.}~\bibnamefont {Yi}},\ and\
		\bibinfo {author} {\bibfnamefont {B.}~\bibnamefont {Yan}},\ }\bibfield
	{title} {\bibinfo {title} {{Dynamic signatures of {n}on-{H}ermitian skin
				effect and topology in ultracold atoms}},\ }\href
	{https://doi.org/10.1103/PhysRevLett.129.070401} {\bibfield  {journal}
		{\bibinfo  {journal} {Phys. Rev. Lett.}\ }\textbf {\bibinfo {volume} {129}},\
		\bibinfo {pages} {070401} (\bibinfo {year} {2022})}\BibitemShut {NoStop}%
	\bibitem [{\citenamefont {Poli}\ \emph {et~al.}(2015)\citenamefont {Poli},
		\citenamefont {Bellec}, \citenamefont {Kuhl}, \citenamefont {Mortessagne},\
		and\ \citenamefont {Schomerus}}]{Poli2015}%
	\BibitemOpen
	\bibfield  {author} {\bibinfo {author} {\bibfnamefont {C.}~\bibnamefont
			{Poli}}, \bibinfo {author} {\bibfnamefont {M.}~\bibnamefont {Bellec}},
		\bibinfo {author} {\bibfnamefont {U.}~\bibnamefont {Kuhl}}, \bibinfo {author}
		{\bibfnamefont {F.}~\bibnamefont {Mortessagne}},\ and\ \bibinfo {author}
		{\bibfnamefont {H.}~\bibnamefont {Schomerus}},\ }\bibfield  {title} {\bibinfo
		{title} {{Selective enhancement of topologically induced interface states in
				a dielectric resonator chain}},\ }\href {https://doi.org/10.1038/ncomms7710}
	{\bibfield  {journal} {\bibinfo  {journal} {Nat. Commun.}\ }\textbf {\bibinfo
			{volume} {6}},\ \bibinfo {pages} {6710} (\bibinfo {year} {2015})}\BibitemShut
	{NoStop}%
	\bibitem [{\citenamefont {Bandres}\ \emph {et~al.}(2018)\citenamefont
		{Bandres}, \citenamefont {Wittek}, \citenamefont {Harari}, \citenamefont
		{Parto}, \citenamefont {Ren}, \citenamefont {Segev}, \citenamefont
		{Christodoulides},\ and\ \citenamefont
		{Khajavikhan}}]{doi:10.1126/science.aar4005}%
	\BibitemOpen
	\bibfield  {author} {\bibinfo {author} {\bibfnamefont {M.~A.}\ \bibnamefont
			{Bandres}}, \bibinfo {author} {\bibfnamefont {S.}~\bibnamefont {Wittek}},
		\bibinfo {author} {\bibfnamefont {G.}~\bibnamefont {Harari}}, \bibinfo
		{author} {\bibfnamefont {M.}~\bibnamefont {Parto}}, \bibinfo {author}
		{\bibfnamefont {J.}~\bibnamefont {Ren}}, \bibinfo {author} {\bibfnamefont
			{M.}~\bibnamefont {Segev}}, \bibinfo {author} {\bibfnamefont {D.~N.}\
			\bibnamefont {Christodoulides}},\ and\ \bibinfo {author} {\bibfnamefont
			{M.}~\bibnamefont {Khajavikhan}},\ }\bibfield  {title} {\bibinfo {title}
		{{Topological insulator laser: Experiments}},\ }\href
	{https://doi.org/10.1126/science.aar4005} {\bibfield  {journal} {\bibinfo
			{journal} {Science}\ }\textbf {\bibinfo {volume} {359}},\ \bibinfo {pages}
		{eaar4005} (\bibinfo {year} {2018})}\BibitemShut {NoStop}%
	\bibitem [{\citenamefont {Parto}\ \emph {et~al.}(2018)\citenamefont {Parto},
		\citenamefont {Wittek}, \citenamefont {Hodaei}, \citenamefont {Harari},
		\citenamefont {Bandres}, \citenamefont {Ren}, \citenamefont {Rechtsman},
		\citenamefont {Segev}, \citenamefont {Christodoulides},\ and\ \citenamefont
		{Khajavikhan}}]{PhysRevLett.120.113901}%
	\BibitemOpen
	\bibfield  {author} {\bibinfo {author} {\bibfnamefont {M.}~\bibnamefont
			{Parto}}, \bibinfo {author} {\bibfnamefont {S.}~\bibnamefont {Wittek}},
		\bibinfo {author} {\bibfnamefont {H.}~\bibnamefont {Hodaei}}, \bibinfo
		{author} {\bibfnamefont {G.}~\bibnamefont {Harari}}, \bibinfo {author}
		{\bibfnamefont {M.~A.}\ \bibnamefont {Bandres}}, \bibinfo {author}
		{\bibfnamefont {J.}~\bibnamefont {Ren}}, \bibinfo {author} {\bibfnamefont
			{M.~C.}\ \bibnamefont {Rechtsman}}, \bibinfo {author} {\bibfnamefont
			{M.}~\bibnamefont {Segev}}, \bibinfo {author} {\bibfnamefont {D.~N.}\
			\bibnamefont {Christodoulides}},\ and\ \bibinfo {author} {\bibfnamefont
			{M.}~\bibnamefont {Khajavikhan}},\ }\bibfield  {title} {\bibinfo {title}
		{{Edge-mode lasing in {1D} topological active arrays}},\ }\href
	{https://doi.org/10.1103/PhysRevLett.120.113901} {\bibfield  {journal}
		{\bibinfo  {journal} {Phys. Rev. Lett.}\ }\textbf {\bibinfo {volume} {120}},\
		\bibinfo {pages} {113901} (\bibinfo {year} {2018})}\BibitemShut {NoStop}%
	\bibitem [{\citenamefont {Xiao}\ \emph {et~al.}(2017)\citenamefont {Xiao},
		\citenamefont {Zhan}, \citenamefont {Bian}, \citenamefont {Wang},
		\citenamefont {Zhang}, \citenamefont {Wang}, \citenamefont {Li},
		\citenamefont {Mochizuki}, \citenamefont {Kim}, \citenamefont {Kawakami},
		\citenamefont {Yi}, \citenamefont {Obuse}, \citenamefont {Sanders},\ and\
		\citenamefont {Xue}}]{xiao2017observation}%
	\BibitemOpen
	\bibfield  {author} {\bibinfo {author} {\bibfnamefont {L.}~\bibnamefont
			{Xiao}}, \bibinfo {author} {\bibfnamefont {X.}~\bibnamefont {Zhan}}, \bibinfo
		{author} {\bibfnamefont {Z.~H.}\ \bibnamefont {Bian}}, \bibinfo {author}
		{\bibfnamefont {K.~K.}\ \bibnamefont {Wang}}, \bibinfo {author}
		{\bibfnamefont {X.}~\bibnamefont {Zhang}}, \bibinfo {author} {\bibfnamefont
			{X.~P.}\ \bibnamefont {Wang}}, \bibinfo {author} {\bibfnamefont
			{J.}~\bibnamefont {Li}}, \bibinfo {author} {\bibfnamefont {K.}~\bibnamefont
			{Mochizuki}}, \bibinfo {author} {\bibfnamefont {D.}~\bibnamefont {Kim}},
		\bibinfo {author} {\bibfnamefont {N.}~\bibnamefont {Kawakami}}, \bibinfo
		{author} {\bibfnamefont {W.}~\bibnamefont {Yi}}, \bibinfo {author}
		{\bibfnamefont {H.}~\bibnamefont {Obuse}}, \bibinfo {author} {\bibfnamefont
			{B.~C.}\ \bibnamefont {Sanders}},\ and\ \bibinfo {author} {\bibfnamefont
			{P.}~\bibnamefont {Xue}},\ }\bibfield  {title} {\bibinfo {title}
		{{Observation of topological edge states in {p}arity--{t}ime-symmetric
				quantum walks}},\ }\href {https://doi.org/10.1038/nphys4204} {\bibfield
		{journal} {\bibinfo  {journal} {Nat. Phys.}\ }\textbf {\bibinfo {volume}
			{13}},\ \bibinfo {pages} {1117} (\bibinfo {year} {2017})}\BibitemShut
	{NoStop}%
	\bibitem [{\citenamefont {Xiao}\ \emph {et~al.}(2026)\citenamefont {Xiao},
		\citenamefont {Sarkar}, \citenamefont {Wang}, \citenamefont {Bayat},\ and\
		\citenamefont {Xue}}]{6gql-zgkb}%
	\BibitemOpen
	\bibfield  {author} {\bibinfo {author} {\bibfnamefont {L.}~\bibnamefont
			{Xiao}}, \bibinfo {author} {\bibfnamefont {S.}~\bibnamefont {Sarkar}},
		\bibinfo {author} {\bibfnamefont {K.}~\bibnamefont {Wang}}, \bibinfo {author}
		{\bibfnamefont {A.}~\bibnamefont {Bayat}},\ and\ \bibinfo {author}
		{\bibfnamefont {P.}~\bibnamefont {Xue}},\ }\bibfield  {title} {\bibinfo
		{title} {{Observation of criticality-enhanced quantum sensing in nonunitary
				quantum walks}},\ }\href {https://doi.org/10.1103/6gql-zgkb} {\bibfield
		{journal} {\bibinfo  {journal} {Phys. Rev. Lett.}\ }\textbf {\bibinfo
			{volume} {136}},\ \bibinfo {pages} {060802} (\bibinfo {year}
		{2026})}\BibitemShut {NoStop}%
	\bibitem [{\citenamefont {Lin}\ \emph {et~al.}(2023)\citenamefont {Lin},
		\citenamefont {Tai}, \citenamefont {Li},\ and\ \citenamefont
		{Lee}}]{lin2023topological}%
	\BibitemOpen
	\bibfield  {author} {\bibinfo {author} {\bibfnamefont {R.}~\bibnamefont
			{Lin}}, \bibinfo {author} {\bibfnamefont {T.}~\bibnamefont {Tai}}, \bibinfo
		{author} {\bibfnamefont {L.}~\bibnamefont {Li}},\ and\ \bibinfo {author}
		{\bibfnamefont {C.~H.}\ \bibnamefont {Lee}},\ }\bibfield  {title} {\bibinfo
		{title} {{Topological {n}on-{H}ermitian skin effect}},\ }\href
	{https://doi.org/10.1007/s11467-023-1309-z} {\bibfield  {journal} {\bibinfo
			{journal} {Front. Phys.}\ }\textbf {\bibinfo {volume} {18}},\ \bibinfo
		{pages} {53605} (\bibinfo {year} {2023})}\BibitemShut {NoStop}%
	\bibitem [{\citenamefont {Borgnia}\ \emph {et~al.}(2020)\citenamefont
		{Borgnia}, \citenamefont {Kruchkov},\ and\ \citenamefont
		{Slager}}]{PhysRevLett.124.056802}%
	\BibitemOpen
	\bibfield  {author} {\bibinfo {author} {\bibfnamefont {D.~S.}\ \bibnamefont
			{Borgnia}}, \bibinfo {author} {\bibfnamefont {A.~J.}\ \bibnamefont
			{Kruchkov}},\ and\ \bibinfo {author} {\bibfnamefont {R.~J.}\ \bibnamefont
			{Slager}},\ }\bibfield  {title} {\bibinfo {title} {{{N}on-{H}ermitian
				boundary modes and topology}},\ }\href
	{https://doi.org/10.1103/PhysRevLett.124.056802} {\bibfield  {journal}
		{\bibinfo  {journal} {Phys. Rev. Lett.}\ }\textbf {\bibinfo {volume} {124}},\
		\bibinfo {pages} {056802} (\bibinfo {year} {2020})}\BibitemShut {NoStop}%
	\bibitem [{\citenamefont {Lee}\ and\ \citenamefont
		{Thomale}(2019)}]{PhysRevB.99.201103}%
	\BibitemOpen
	\bibfield  {author} {\bibinfo {author} {\bibfnamefont {C.~H.}\ \bibnamefont
			{Lee}}\ and\ \bibinfo {author} {\bibfnamefont {R.}~\bibnamefont {Thomale}},\
	}\bibfield  {title} {\bibinfo {title} {{Anatomy of skin modes and topology in
				{n}on-{H}ermitian systems}},\ }\href
	{https://doi.org/10.1103/PhysRevB.99.201103} {\bibfield  {journal} {\bibinfo
			{journal} {Phys. Rev. B}\ }\textbf {\bibinfo {volume} {99}},\ \bibinfo
		{pages} {201103} (\bibinfo {year} {2019})}\BibitemShut {NoStop}%
	\bibitem [{\citenamefont {Li}\ \emph {et~al.}(2019)\citenamefont {Li},
		\citenamefont {Lee},\ and\ \citenamefont {Gong}}]{PhysRevB.100.075403}%
	\BibitemOpen
	\bibfield  {author} {\bibinfo {author} {\bibfnamefont {L.}~\bibnamefont
			{Li}}, \bibinfo {author} {\bibfnamefont {C.~H.}\ \bibnamefont {Lee}},\ and\
		\bibinfo {author} {\bibfnamefont {J.}~\bibnamefont {Gong}},\ }\bibfield
	{title} {\bibinfo {title} {{Geometric characterization of {n}on-{H}ermitian
				topological systems through the singularity ring in pseudospin vector
				space}},\ }\href {https://doi.org/10.1103/PhysRevB.100.075403} {\bibfield
		{journal} {\bibinfo  {journal} {Phys. Rev. B}\ }\textbf {\bibinfo {volume}
			{100}},\ \bibinfo {pages} {075403} (\bibinfo {year} {2019})}\BibitemShut
	{NoStop}%
	\bibitem [{\citenamefont {Okuma}\ \emph {et~al.}(2020)\citenamefont {Okuma},
		\citenamefont {Kawabata}, \citenamefont {Shiozaki},\ and\ \citenamefont
		{Sato}}]{PhysRevLett.124.086801}%
	\BibitemOpen
	\bibfield  {author} {\bibinfo {author} {\bibfnamefont {N.}~\bibnamefont
			{Okuma}}, \bibinfo {author} {\bibfnamefont {K.}~\bibnamefont {Kawabata}},
		\bibinfo {author} {\bibfnamefont {K.}~\bibnamefont {Shiozaki}},\ and\
		\bibinfo {author} {\bibfnamefont {M.}~\bibnamefont {Sato}},\ }\bibfield
	{title} {\bibinfo {title} {{Topological origin of {n}on-{H}ermitian skin
				effects}},\ }\href {https://doi.org/10.1103/PhysRevLett.124.086801}
	{\bibfield  {journal} {\bibinfo  {journal} {Phys. Rev. Lett.}\ }\textbf
		{\bibinfo {volume} {124}},\ \bibinfo {pages} {086801} (\bibinfo {year}
		{2020})}\BibitemShut {NoStop}%
	\bibitem [{\citenamefont {Gong}\ \emph {et~al.}(2018)\citenamefont {Gong},
		\citenamefont {Ashida}, \citenamefont {Kawabata}, \citenamefont {Takasan},
		\citenamefont {Higashikawa},\ and\ \citenamefont {Ueda}}]{PhysRevX.8.031079}%
	\BibitemOpen
	\bibfield  {author} {\bibinfo {author} {\bibfnamefont {Z.}~\bibnamefont
			{Gong}}, \bibinfo {author} {\bibfnamefont {Y.}~\bibnamefont {Ashida}},
		\bibinfo {author} {\bibfnamefont {K.}~\bibnamefont {Kawabata}}, \bibinfo
		{author} {\bibfnamefont {K.}~\bibnamefont {Takasan}}, \bibinfo {author}
		{\bibfnamefont {S.}~\bibnamefont {Higashikawa}},\ and\ \bibinfo {author}
		{\bibfnamefont {M.}~\bibnamefont {Ueda}},\ }\bibfield  {title} {\bibinfo
		{title} {{Topological phases of {n}on-{H}ermitian systems}},\ }\href
	{https://doi.org/10.1103/PhysRevX.8.031079} {\bibfield  {journal} {\bibinfo
			{journal} {Phys. Rev. X}\ }\textbf {\bibinfo {volume} {8}},\ \bibinfo {pages}
		{031079} (\bibinfo {year} {2018})}\BibitemShut {NoStop}%
	\bibitem [{\citenamefont {Zhang}\ \emph {et~al.}(2020)\citenamefont {Zhang},
		\citenamefont {Yang},\ and\ \citenamefont {Fang}}]{PhysRevLett.125.126402}%
	\BibitemOpen
	\bibfield  {author} {\bibinfo {author} {\bibfnamefont {K.}~\bibnamefont
			{Zhang}}, \bibinfo {author} {\bibfnamefont {Z.}~\bibnamefont {Yang}},\ and\
		\bibinfo {author} {\bibfnamefont {C.}~\bibnamefont {Fang}},\ }\bibfield
	{title} {\bibinfo {title} {{Correspondence between winding numbers and skin
				modes in {n}on-{H}ermitian systems}},\ }\href
	{https://doi.org/10.1103/PhysRevLett.125.126402} {\bibfield  {journal}
		{\bibinfo  {journal} {Phys. Rev. Lett.}\ }\textbf {\bibinfo {volume} {125}},\
		\bibinfo {pages} {126402} (\bibinfo {year} {2020})}\BibitemShut {NoStop}%
	\bibitem [{\citenamefont {Kawabata}\ \emph {et~al.}(2023)\citenamefont
		{Kawabata}, \citenamefont {Numasawa},\ and\ \citenamefont
		{Ryu}}]{PhysRevX.13.021007}%
	\BibitemOpen
	\bibfield  {author} {\bibinfo {author} {\bibfnamefont {K.}~\bibnamefont
			{Kawabata}}, \bibinfo {author} {\bibfnamefont {T.}~\bibnamefont {Numasawa}},\
		and\ \bibinfo {author} {\bibfnamefont {S.}~\bibnamefont {Ryu}},\ }\bibfield
	{title} {\bibinfo {title} {{Entanglement phase transition induced by the
				{n}on-{H}ermitian skin effect}},\ }\href
	{https://doi.org/10.1103/PhysRevX.13.021007} {\bibfield  {journal} {\bibinfo
			{journal} {Phys. Rev. X}\ }\textbf {\bibinfo {volume} {13}},\ \bibinfo
		{pages} {021007} (\bibinfo {year} {2023})}\BibitemShut {NoStop}%
	\bibitem [{\citenamefont {Lee}(2016)}]{PhysRevLett.116.133903}%
	\BibitemOpen
	\bibfield  {author} {\bibinfo {author} {\bibfnamefont {T.~E.}\ \bibnamefont
			{Lee}},\ }\bibfield  {title} {\bibinfo {title} {{Anomalous edge state in a
				{n}on-{H}ermitian lattice}},\ }\href
	{https://doi.org/10.1103/PhysRevLett.116.133903} {\bibfield  {journal}
		{\bibinfo  {journal} {Phys. Rev. Lett.}\ }\textbf {\bibinfo {volume} {116}},\
		\bibinfo {pages} {133903} (\bibinfo {year} {2016})}\BibitemShut {NoStop}%
	\bibitem [{\citenamefont {Lee}\ \emph {et~al.}(2019)\citenamefont {Lee},
		\citenamefont {Li},\ and\ \citenamefont {Gong}}]{PhysRevLett.123.016805}%
	\BibitemOpen
	\bibfield  {author} {\bibinfo {author} {\bibfnamefont {C.~H.}\ \bibnamefont
			{Lee}}, \bibinfo {author} {\bibfnamefont {L.}~\bibnamefont {Li}},\ and\
		\bibinfo {author} {\bibfnamefont {J.}~\bibnamefont {Gong}},\ }\bibfield
	{title} {\bibinfo {title} {{Hybrid higher-order skin-topological modes in
				nonreciprocal systems}},\ }\href
	{https://doi.org/10.1103/PhysRevLett.123.016805} {\bibfield  {journal}
		{\bibinfo  {journal} {Phys. Rev. Lett.}\ }\textbf {\bibinfo {volume} {123}},\
		\bibinfo {pages} {016805} (\bibinfo {year} {2019})}\BibitemShut {NoStop}%
	\bibitem [{\citenamefont {Aquino}\ \emph {et~al.}(2023)\citenamefont {Aquino},
		\citenamefont {Lopes},\ and\ \citenamefont {Barci}}]{PhysRevB.107.035424}%
	\BibitemOpen
	\bibfield  {author} {\bibinfo {author} {\bibfnamefont {R.}~\bibnamefont
			{Aquino}}, \bibinfo {author} {\bibfnamefont {N.}~\bibnamefont {Lopes}},\ and\
		\bibinfo {author} {\bibfnamefont {D.~G.}\ \bibnamefont {Barci}},\ }\bibfield
	{title} {\bibinfo {title} {{Critical and noncritical {n}on-{H}ermitian
				topological phase transitions in one-dimensional chains}},\ }\href
	{https://doi.org/10.1103/PhysRevB.107.035424} {\bibfield  {journal} {\bibinfo
			{journal} {Phys. Rev. B}\ }\textbf {\bibinfo {volume} {107}},\ \bibinfo
		{pages} {035424} (\bibinfo {year} {2023})}\BibitemShut {NoStop}%
	\bibitem [{\citenamefont {Kunst}\ \emph {et~al.}(2018)\citenamefont {Kunst},
		\citenamefont {Edvardsson}, \citenamefont {Budich},\ and\ \citenamefont
		{Bergholtz}}]{PhysRevLett.121.026808}%
	\BibitemOpen
	\bibfield  {author} {\bibinfo {author} {\bibfnamefont {F.~K.}\ \bibnamefont
			{Kunst}}, \bibinfo {author} {\bibfnamefont {E.}~\bibnamefont {Edvardsson}},
		\bibinfo {author} {\bibfnamefont {J.~C.}\ \bibnamefont {Budich}},\ and\
		\bibinfo {author} {\bibfnamefont {E.~J.}\ \bibnamefont {Bergholtz}},\
	}\bibfield  {title} {\bibinfo {title} {{Biorthogonal bulk-boundary
				correspondence in {n}on-{H}ermitian systems}},\ }\href
	{https://doi.org/10.1103/PhysRevLett.121.026808} {\bibfield  {journal}
		{\bibinfo  {journal} {Phys. Rev. Lett.}\ }\textbf {\bibinfo {volume} {121}},\
		\bibinfo {pages} {026808} (\bibinfo {year} {2018})}\BibitemShut {NoStop}%
	\bibitem [{\citenamefont {Yi}\ and\ \citenamefont
		{Yang}(2020)}]{PhysRevLett.125.186802}%
	\BibitemOpen
	\bibfield  {author} {\bibinfo {author} {\bibfnamefont {Y.}~\bibnamefont
			{Yi}}\ and\ \bibinfo {author} {\bibfnamefont {Z.}~\bibnamefont {Yang}},\
	}\bibfield  {title} {\bibinfo {title} {{{N}on-{H}ermitian skin modes induced
				by on-site dissipations and chiral tunneling effect}},\ }\href
	{https://doi.org/10.1103/PhysRevLett.125.186802} {\bibfield  {journal}
		{\bibinfo  {journal} {Phys. Rev. Lett.}\ }\textbf {\bibinfo {volume} {125}},\
		\bibinfo {pages} {186802} (\bibinfo {year} {2020})}\BibitemShut {NoStop}%
	\bibitem [{\citenamefont {Song}\ \emph {et~al.}(2019)\citenamefont {Song},
		\citenamefont {Yao},\ and\ \citenamefont {Wang}}]{PhysRevLett.123.246801}%
	\BibitemOpen
	\bibfield  {author} {\bibinfo {author} {\bibfnamefont {F.}~\bibnamefont
			{Song}}, \bibinfo {author} {\bibfnamefont {S.}~\bibnamefont {Yao}},\ and\
		\bibinfo {author} {\bibfnamefont {Z.}~\bibnamefont {Wang}},\ }\bibfield
	{title} {\bibinfo {title} {{{N}on-{H}ermitian topological invariants in real
				space}},\ }\href {https://doi.org/10.1103/PhysRevLett.123.246801} {\bibfield
		{journal} {\bibinfo  {journal} {Phys. Rev. Lett.}\ }\textbf {\bibinfo
			{volume} {123}},\ \bibinfo {pages} {246801} (\bibinfo {year}
		{2019})}\BibitemShut {NoStop}%
	\bibitem [{\citenamefont {Longhi}(2022)}]{PhysRevLett.128.157601}%
	\BibitemOpen
	\bibfield  {author} {\bibinfo {author} {\bibfnamefont {S.}~\bibnamefont
			{Longhi}},\ }\bibfield  {title} {\bibinfo {title} {{Self-healing of
				{n}on-{H}ermitian topological skin modes}},\ }\href
	{https://doi.org/10.1103/PhysRevLett.128.157601} {\bibfield  {journal}
		{\bibinfo  {journal} {Phys. Rev. Lett.}\ }\textbf {\bibinfo {volume} {128}},\
		\bibinfo {pages} {157601} (\bibinfo {year} {2022})}\BibitemShut {NoStop}%
		\bibitem [{\citenamefont {Nakamura}\ \emph {et~al.}(2024)\citenamefont
		{Nakamura}, \citenamefont {Bessho},\ and\ \citenamefont
		{Sato}}]{PhysRevLett.132.136401}%
	\BibitemOpen
	\bibfield  {author} {\bibinfo {author} {\bibfnamefont {D.}~\bibnamefont
			{Nakamura}}, \bibinfo {author} {\bibfnamefont {T.}~\bibnamefont {Bessho}},\
		and\ \bibinfo {author} {\bibfnamefont {M.}~\bibnamefont {Sato}},\ }\bibfield
	{title} {\bibinfo {title} {{Bulk-boundary correspondence in point-gap
				topological phases}},\ }\href
	{https://doi.org/10.1103/PhysRevLett.132.136401} {\bibfield  {journal}
		{\bibinfo  {journal} {Phys. Rev. Lett.}\ }\textbf {\bibinfo {volume} {132}},\
		\bibinfo {pages} {136401} (\bibinfo {year} {2024})}\BibitemShut {NoStop}%
	\bibitem [{\citenamefont {Kawabata}\ \emph {et~al.}(2019)\citenamefont
		{Kawabata}, \citenamefont {Shiozaki}, \citenamefont {Ueda},\ and\
		\citenamefont {Sato}}]{PhysRevX.9.041015}%
	\BibitemOpen
	\bibfield  {author} {\bibinfo {author} {\bibfnamefont {K.}~\bibnamefont
			{Kawabata}}, \bibinfo {author} {\bibfnamefont {K.}~\bibnamefont {Shiozaki}},
		\bibinfo {author} {\bibfnamefont {M.}~\bibnamefont {Ueda}},\ and\ \bibinfo
		{author} {\bibfnamefont {M.}~\bibnamefont {Sato}},\ }\bibfield  {title}
	{\bibinfo {title} {{Symmetry and topology in {n}on-{H}ermitian physics}},\
	}\href {https://doi.org/10.1103/PhysRevX.9.041015} {\bibfield  {journal}
		{\bibinfo  {journal} {Phys. Rev. X}\ }\textbf {\bibinfo {volume} {9}},\
		\bibinfo {pages} {041015} (\bibinfo {year} {2019})}\BibitemShut {NoStop}%
	\bibitem [{\citenamefont {Yang}\ \emph {et~al.}(2020)\citenamefont {Yang},
		\citenamefont {Zhang}, \citenamefont {Fang},\ and\ \citenamefont
		{Hu}}]{PhysRevLett.125.226402}%
	\BibitemOpen
	\bibfield  {author} {\bibinfo {author} {\bibfnamefont {Z.}~\bibnamefont
			{Yang}}, \bibinfo {author} {\bibfnamefont {K.}~\bibnamefont {Zhang}},
		\bibinfo {author} {\bibfnamefont {C.}~\bibnamefont {Fang}},\ and\ \bibinfo
		{author} {\bibfnamefont {J.}~\bibnamefont {Hu}},\ }\bibfield  {title}
	{\bibinfo {title} {{{N}on-{H}ermitian bulk-boundary correspondence and
				auxiliary generalized {B}rillouin zone theory}},\ }\href
	{https://doi.org/10.1103/PhysRevLett.125.226402} {\bibfield  {journal}
		{\bibinfo  {journal} {Phys. Rev. Lett.}\ }\textbf {\bibinfo {volume} {125}},\
		\bibinfo {pages} {226402} (\bibinfo {year} {2020})}\BibitemShut {NoStop}%
	\bibitem [{\citenamefont {Li}\ \emph {et~al.}(2025)\citenamefont {Li},
		\citenamefont {Li},\ and\ \citenamefont {Xu}}]{li2025size}%
	\BibitemOpen
	\bibfield  {author} {\bibinfo {author} {\bibfnamefont {Y.}~\bibnamefont
			{Li}}, \bibinfo {author} {\bibfnamefont {L.}~\bibnamefont {Li}},\ and\
		\bibinfo {author} {\bibfnamefont {Z.}~\bibnamefont {Xu}},\ }\bibfield
	{title} {\bibinfo {title} {{Size-dependent skin effect transitions in weakly
				coupled nonreciprocal chains}},\ }\href {https://doi.org/10.1103/bmq5-7tf6}
	{\bibfield  {journal} {\bibinfo  {journal} {Phys. Rev. B}\ }\textbf {\bibinfo
			{volume} {112}},\ \bibinfo {pages} {235122} (\bibinfo {year}
		{2025})}\BibitemShut {NoStop}%
	\bibitem [{\citenamefont {Yao}\ and\ \citenamefont
		{Wang}(2018)}]{PhysRevLett.121.086803}%
	\BibitemOpen
	\bibfield  {author} {\bibinfo {author} {\bibfnamefont {S.}~\bibnamefont
			{Yao}}\ and\ \bibinfo {author} {\bibfnamefont {Z.}~\bibnamefont {Wang}},\
	}\bibfield  {title} {\bibinfo {title} {{Edge states and topological
				invariants of {n}on-{H}ermitian systems}},\ }\href
	{https://doi.org/10.1103/PhysRevLett.121.086803} {\bibfield  {journal}
		{\bibinfo  {journal} {Phys. Rev. Lett.}\ }\textbf {\bibinfo {volume} {121}},\
		\bibinfo {pages} {086803} (\bibinfo {year} {2018})}\BibitemShut {NoStop}%
	\bibitem [{\citenamefont {Yao}\ \emph {et~al.}(2018)\citenamefont {Yao},
		\citenamefont {Song},\ and\ \citenamefont {Wang}}]{PhysRevLett.121.136802}%
	\BibitemOpen
	\bibfield  {author} {\bibinfo {author} {\bibfnamefont {S.}~\bibnamefont
			{Yao}}, \bibinfo {author} {\bibfnamefont {F.}~\bibnamefont {Song}},\ and\
		\bibinfo {author} {\bibfnamefont {Z.}~\bibnamefont {Wang}},\ }\bibfield
	{title} {\bibinfo {title} {{{N}on-{H}ermitian {C}hern bands}},\ }\href
	{https://doi.org/10.1103/PhysRevLett.121.136802} {\bibfield  {journal}
		{\bibinfo  {journal} {Phys. Rev. Lett.}\ }\textbf {\bibinfo {volume} {121}},\
		\bibinfo {pages} {136802} (\bibinfo {year} {2018})}\BibitemShut {NoStop}%
	\bibitem [{\citenamefont {Yokomizo}\ and\ \citenamefont
		{Murakami}(2019)}]{PhysRevLett.123.066404}%
	\BibitemOpen
	\bibfield  {author} {\bibinfo {author} {\bibfnamefont {K.}~\bibnamefont
			{Yokomizo}}\ and\ \bibinfo {author} {\bibfnamefont {S.}~\bibnamefont
			{Murakami}},\ }\bibfield  {title} {\bibinfo {title} {{{N}on-{B}loch band
				theory of {n}on-{H}ermitian systems}},\ }\href
	{https://doi.org/10.1103/PhysRevLett.123.066404} {\bibfield  {journal}
		{\bibinfo  {journal} {Phys. Rev. Lett.}\ }\textbf {\bibinfo {volume} {123}},\
		\bibinfo {pages} {066404} (\bibinfo {year} {2019})}\BibitemShut {NoStop}%
	\bibitem [{\citenamefont {Yokomizo}\ and\ \citenamefont
		{Murakami}(2020)}]{10.1093/ptep/ptaa140}%
	\BibitemOpen
	\bibfield  {author} {\bibinfo {author} {\bibfnamefont {K.}~\bibnamefont
			{Yokomizo}}\ and\ \bibinfo {author} {\bibfnamefont {S.}~\bibnamefont
			{Murakami}},\ }\bibfield  {title} {\bibinfo {title} {{{N}on-{B}loch band
				theory and bulk--edge correspondence in {n}on-{H}ermitian systems}},\ }\href
	{https://doi.org/10.1093/ptep/ptaa140} {\bibfield  {journal} {\bibinfo
			{journal} {Prog. Theor. Exp. Phys.}\ }\textbf {\bibinfo {volume} {2020}},\
		\bibinfo {pages} {12A102} (\bibinfo {year} {2020})}\BibitemShut {NoStop}%
	\bibitem [{\citenamefont {Xu}\ \emph {et~al.}(2021)\citenamefont {Xu},
		\citenamefont {Zhang}, \citenamefont {Luo}, \citenamefont {Yu}, \citenamefont
		{Li},\ and\ \citenamefont {Zhang}}]{PhysRevB.103.125411}%
	\BibitemOpen
	\bibfield  {author} {\bibinfo {author} {\bibfnamefont {K.}~\bibnamefont
			{Xu}}, \bibinfo {author} {\bibfnamefont {X.}~\bibnamefont {Zhang}}, \bibinfo
		{author} {\bibfnamefont {K.}~\bibnamefont {Luo}}, \bibinfo {author}
		{\bibfnamefont {R.}~\bibnamefont {Yu}}, \bibinfo {author} {\bibfnamefont
			{D.}~\bibnamefont {Li}},\ and\ \bibinfo {author} {\bibfnamefont
			{H.}~\bibnamefont {Zhang}},\ }\bibfield  {title} {\bibinfo {title}
		{{Coexistence of topological edge states and skin effects in the
				{n}on-{H}ermitian {S}u-{S}chrieffer-{H}eeger model with long-range
				nonreciprocal hopping in topoelectric realizations}},\ }\href
	{https://doi.org/10.1103/PhysRevB.103.125411} {\bibfield  {journal} {\bibinfo
			{journal} {Phys. Rev. B}\ }\textbf {\bibinfo {volume} {103}},\ \bibinfo
		{pages} {125411} (\bibinfo {year} {2021})}\BibitemShut {NoStop}%
	\bibitem [{\citenamefont {Yang}\ and\ \citenamefont
		{Bergholtz}(2025)}]{PhysRevResearch.7.023233}%
	\BibitemOpen
	\bibfield  {author} {\bibinfo {author} {\bibfnamefont {F.}~\bibnamefont
			{Yang}}\ and\ \bibinfo {author} {\bibfnamefont {E.~J.}\ \bibnamefont
			{Bergholtz}},\ }\bibfield  {title} {\bibinfo {title} {{Anatomy of
				higher-order {n}on-{H}ermitian skin and boundary modes}},\ }\href
	{https://doi.org/10.1103/PhysRevResearch.7.023233} {\bibfield  {journal}
		{\bibinfo  {journal} {Phys. Rev. Res.}\ }\textbf {\bibinfo {volume} {7}},\
		\bibinfo {pages} {023233} (\bibinfo {year} {2025})}\BibitemShut {NoStop}%
	\bibitem [{\citenamefont {Hamanaka}\ \emph {et~al.}(2024)\citenamefont
		{Hamanaka}, \citenamefont {Yoshida},\ and\ \citenamefont
		{Kawabata}}]{PhysRevLett.133.266604}%
	\BibitemOpen
	\bibfield  {author} {\bibinfo {author} {\bibfnamefont {S.}~\bibnamefont
			{Hamanaka}}, \bibinfo {author} {\bibfnamefont {T.}~\bibnamefont {Yoshida}},\
		and\ \bibinfo {author} {\bibfnamefont {K.}~\bibnamefont {Kawabata}},\
	}\bibfield  {title} {\bibinfo {title} {{{N}on-{H}ermitian topology in
				{H}ermitian topological matter}},\ }\href
	{https://doi.org/10.1103/PhysRevLett.133.266604} {\bibfield  {journal}
		{\bibinfo  {journal} {Phys. Rev. Lett.}\ }\textbf {\bibinfo {volume} {133}},\
		\bibinfo {pages} {266604} (\bibinfo {year} {2024})}\BibitemShut {NoStop}%
	\bibitem [{\citenamefont {Wan}\ and\ \citenamefont
		{L{\"u}}(2023)}]{PhysRevLett.130.203605}%
	\BibitemOpen
	\bibfield  {author} {\bibinfo {author} {\bibfnamefont {L.-L.}\ \bibnamefont
			{Wan}}\ and\ \bibinfo {author} {\bibfnamefont {X.-Y.}\ \bibnamefont
			{L{\"u}}},\ }\bibfield  {title} {\bibinfo {title} {{Quantum-squeezing-induced
				point-gap topology and skin effect}},\ }\href
	{https://doi.org/10.1103/PhysRevLett.130.203605} {\bibfield  {journal}
		{\bibinfo  {journal} {Phys. Rev. Lett.}\ }\textbf {\bibinfo {volume} {130}},\
		\bibinfo {pages} {203605} (\bibinfo {year} {2023})}\BibitemShut {NoStop}%
	\bibitem [{\citenamefont {Schindler}\ \emph {et~al.}(2023)\citenamefont
		{Schindler}, \citenamefont {Gu}, \citenamefont {Lian},\ and\ \citenamefont
		{Kawabata}}]{PRXQuantum.4.030315}%
	\BibitemOpen
	\bibfield  {author} {\bibinfo {author} {\bibfnamefont {F.}~\bibnamefont
			{Schindler}}, \bibinfo {author} {\bibfnamefont {K.}~\bibnamefont {Gu}},
		\bibinfo {author} {\bibfnamefont {B.}~\bibnamefont {Lian}},\ and\ \bibinfo
		{author} {\bibfnamefont {K.}~\bibnamefont {Kawabata}},\ }\bibfield  {title}
	{\bibinfo {title} {{Hermitian bulk -- {n}on-{H}ermitian boundary
				correspondence}},\ }\href {https://doi.org/10.1103/PRXQuantum.4.030315}
	{\bibfield  {journal} {\bibinfo  {journal} {PRX Quantum}\ }\textbf {\bibinfo
			{volume} {4}},\ \bibinfo {pages} {030315} (\bibinfo {year}
		{2023})}\BibitemShut {NoStop}%
	\bibitem [{\citenamefont {Liu}\ \emph {et~al.}(2023)\citenamefont {Liu},
		\citenamefont {Wang}, \citenamefont {Cheng}, \citenamefont {Hu},
		\citenamefont {Wang}, \citenamefont {Xue}, \citenamefont {Zhang},\ and\
		\citenamefont {Luo}}]{liu2023simultaneous}%
	\BibitemOpen
	\bibfield  {author} {\bibinfo {author} {\bibfnamefont {D.}~\bibnamefont
			{Liu}}, \bibinfo {author} {\bibfnamefont {Z.}~\bibnamefont {Wang}}, \bibinfo
		{author} {\bibfnamefont {Z.}~\bibnamefont {Cheng}}, \bibinfo {author}
		{\bibfnamefont {H.}~\bibnamefont {Hu}}, \bibinfo {author} {\bibfnamefont
			{Q.}~\bibnamefont {Wang}}, \bibinfo {author} {\bibfnamefont {H.}~\bibnamefont
			{Xue}}, \bibinfo {author} {\bibfnamefont {B.}~\bibnamefont {Zhang}},\ and\
		\bibinfo {author} {\bibfnamefont {Y.}~\bibnamefont {Luo}},\ }\bibfield
	{title} {\bibinfo {title} {{Simultaneous manipulation of line-gap and
				point-gap topologies in {n}on-{H}ermitian lattices}},\ }\href
	{https://doi.org/10.1002/lpor.202200371} {\bibfield  {journal} {\bibinfo
			{journal} {Laser Photonics Rev.}\ }\textbf {\bibinfo {volume} {17}},\
		\bibinfo {pages} {2200371} (\bibinfo {year} {2023})}\BibitemShut {NoStop}%
	\bibitem [{\citenamefont {Qin}\ \emph {et~al.}(2023)\citenamefont {Qin},
		\citenamefont {Ma}, \citenamefont {Shen},\ and\ \citenamefont
		{Lee}}]{PhysRevB.107.155430}%
	\BibitemOpen
	\bibfield  {author} {\bibinfo {author} {\bibfnamefont {F.}~\bibnamefont
			{Qin}}, \bibinfo {author} {\bibfnamefont {Y.}~\bibnamefont {Ma}}, \bibinfo
		{author} {\bibfnamefont {R.}~\bibnamefont {Shen}},\ and\ \bibinfo {author}
		{\bibfnamefont {C.~H.}\ \bibnamefont {Lee}},\ }\bibfield  {title} {\bibinfo
		{title} {{Universal competitive spectral scaling from the critical
				{n}on-{H}ermitian skin effect}},\ }\href
	{https://doi.org/10.1103/PhysRevB.107.155430} {\bibfield  {journal} {\bibinfo
			{journal} {Phys. Rev. B}\ }\textbf {\bibinfo {volume} {107}},\ \bibinfo
		{pages} {155430} (\bibinfo {year} {2023})}\BibitemShut {NoStop}%
	\bibitem [{\citenamefont {Helbig}\ \emph {et~al.}(2020)\citenamefont {Helbig},
		\citenamefont {Hofmann}, \citenamefont {Imhof}, \citenamefont {Abdelghany},
		\citenamefont {Kiessling}, \citenamefont {Molenkamp}, \citenamefont {Lee},
		\citenamefont {Szameit}, \citenamefont {Greiter},\ and\ \citenamefont
		{Thomale}}]{helbig2020generalized}%
	\BibitemOpen
	\bibfield  {author} {\bibinfo {author} {\bibfnamefont {T.}~\bibnamefont
			{Helbig}}, \bibinfo {author} {\bibfnamefont {T.}~\bibnamefont {Hofmann}},
		\bibinfo {author} {\bibfnamefont {S.}~\bibnamefont {Imhof}}, \bibinfo
		{author} {\bibfnamefont {M.}~\bibnamefont {Abdelghany}}, \bibinfo {author}
		{\bibfnamefont {T.}~\bibnamefont {Kiessling}}, \bibinfo {author}
		{\bibfnamefont {L.~W.}\ \bibnamefont {Molenkamp}}, \bibinfo {author}
		{\bibfnamefont {C.~H.}\ \bibnamefont {Lee}}, \bibinfo {author} {\bibfnamefont
			{A.}~\bibnamefont {Szameit}}, \bibinfo {author} {\bibfnamefont
			{M.}~\bibnamefont {Greiter}},\ and\ \bibinfo {author} {\bibfnamefont
			{R.}~\bibnamefont {Thomale}},\ }\bibfield  {title} {\bibinfo {title}
		{{Generalized bulk--boundary correspondence in {n}on-{H}ermitian
				topolectrical circuits}},\ }\href {https://doi.org/10.1038/s41567-020-0922-9}
	{\bibfield  {journal} {\bibinfo  {journal} {Nat. Phys.}\ }\textbf {\bibinfo
			{volume} {16}},\ \bibinfo {pages} {747} (\bibinfo {year} {2020})}\BibitemShut
	{NoStop}%
	\bibitem [{\citenamefont {Xiao}\ \emph {et~al.}(2020)\citenamefont {Xiao},
		\citenamefont {Deng}, \citenamefont {Wang}, \citenamefont {Zhu},
		\citenamefont {Wang}, \citenamefont {Yi},\ and\ \citenamefont
		{Xue}}]{Xiao2020}%
	\BibitemOpen
	\bibfield  {author} {\bibinfo {author} {\bibfnamefont {L.}~\bibnamefont
			{Xiao}}, \bibinfo {author} {\bibfnamefont {T.}~\bibnamefont {Deng}}, \bibinfo
		{author} {\bibfnamefont {K.}~\bibnamefont {Wang}}, \bibinfo {author}
		{\bibfnamefont {G.}~\bibnamefont {Zhu}}, \bibinfo {author} {\bibfnamefont
			{Z.}~\bibnamefont {Wang}}, \bibinfo {author} {\bibfnamefont {W.}~\bibnamefont
			{Yi}},\ and\ \bibinfo {author} {\bibfnamefont {P.}~\bibnamefont {Xue}},\
	}\bibfield  {title} {\bibinfo {title} {{{N}on-{H}ermitian bulk--boundary
				correspondence in quantum dynamics}},\ }\href
	{https://doi.org/10.1038/s41567-020-0836-6} {\bibfield  {journal} {\bibinfo
			{journal} {Nat. Phys.}\ }\textbf {\bibinfo {volume} {16}},\ \bibinfo {pages}
		{761} (\bibinfo {year} {2020})}\BibitemShut {NoStop}%
	\bibitem [{\citenamefont {Weidemann}\ \emph {et~al.}(2020)\citenamefont
		{Weidemann}, \citenamefont {Kremer}, \citenamefont {Helbig}, \citenamefont
		{Hofmann}, \citenamefont {Stegmaier}, \citenamefont {Greiter}, \citenamefont
		{Thomale},\ and\ \citenamefont {Szameit}}]{doi:10.1126/science.aaz8727}%
	\BibitemOpen
	\bibfield  {author} {\bibinfo {author} {\bibfnamefont {S.}~\bibnamefont
			{Weidemann}}, \bibinfo {author} {\bibfnamefont {M.}~\bibnamefont {Kremer}},
		\bibinfo {author} {\bibfnamefont {T.}~\bibnamefont {Helbig}}, \bibinfo
		{author} {\bibfnamefont {T.}~\bibnamefont {Hofmann}}, \bibinfo {author}
		{\bibfnamefont {A.}~\bibnamefont {Stegmaier}}, \bibinfo {author}
		{\bibfnamefont {M.}~\bibnamefont {Greiter}}, \bibinfo {author} {\bibfnamefont
			{R.}~\bibnamefont {Thomale}},\ and\ \bibinfo {author} {\bibfnamefont
			{A.}~\bibnamefont {Szameit}},\ }\bibfield  {title} {\bibinfo {title}
		{{Topological funneling of light}},\ }\href
	{https://doi.org/10.1126/science.aaz8727} {\bibfield  {journal} {\bibinfo
			{journal} {Science}\ }\textbf {\bibinfo {volume} {368}},\ \bibinfo {pages}
		{311} (\bibinfo {year} {2020})}\BibitemShut {NoStop}%
	\bibitem [{\citenamefont {Xiao}\ \emph {et~al.}(2021)\citenamefont {Xiao},
		\citenamefont {Deng}, \citenamefont {Wang}, \citenamefont {Wang},
		\citenamefont {Yi},\ and\ \citenamefont {Xue}}]{PhysRevLett.126.230402}%
	\BibitemOpen
	\bibfield  {author} {\bibinfo {author} {\bibfnamefont {L.}~\bibnamefont
			{Xiao}}, \bibinfo {author} {\bibfnamefont {T.}~\bibnamefont {Deng}}, \bibinfo
		{author} {\bibfnamefont {K.}~\bibnamefont {Wang}}, \bibinfo {author}
		{\bibfnamefont {Z.}~\bibnamefont {Wang}}, \bibinfo {author} {\bibfnamefont
			{W.}~\bibnamefont {Yi}},\ and\ \bibinfo {author} {\bibfnamefont
			{P.}~\bibnamefont {Xue}},\ }\bibfield  {title} {\bibinfo {title}
		{{Observation of {n}on-{B}loch {p}arity-{t}ime symmetry and exceptional
				points}},\ }\href {https://doi.org/10.1103/PhysRevLett.126.230402} {\bibfield
		{journal} {\bibinfo  {journal} {Phys. Rev. Lett.}\ }\textbf {\bibinfo
			{volume} {126}},\ \bibinfo {pages} {230402} (\bibinfo {year}
		{2021})}\BibitemShut {NoStop}%
	\bibitem [{\citenamefont {Heiss}(2004)}]{heiss2004exceptional}%
	\BibitemOpen
	\bibfield  {author} {\bibinfo {author} {\bibfnamefont {W.~D.}\ \bibnamefont
			{Heiss}},\ }\bibfield  {title} {\bibinfo {title} {{Exceptional points of
				{n}on-{H}ermitian operators}},\ }\href
	{https://doi.org/10.1088/0305-4470/37/6/034} {\bibfield  {journal} {\bibinfo
			{journal} {J. Phys. A: Math. Gen.}\ }\textbf {\bibinfo {volume} {37}},\
		\bibinfo {pages} {2455} (\bibinfo {year} {2004})}\BibitemShut {NoStop}%
	\bibitem [{\citenamefont {Miri}\ and\ \citenamefont
		{Alu}(2019)}]{miri2019exceptional}%
	\BibitemOpen
	\bibfield  {author} {\bibinfo {author} {\bibfnamefont {M.~A.}\ \bibnamefont
			{Miri}}\ and\ \bibinfo {author} {\bibfnamefont {A.}~\bibnamefont {Alu}},\
	}\bibfield  {title} {\bibinfo {title} {{Exceptional points in optics and
				photonics}},\ }\href {https://doi.org/10.1126/science.aar7709} {\bibfield
		{journal} {\bibinfo  {journal} {Science}\ }\textbf {\bibinfo {volume}
			{363}},\ \bibinfo {pages} {eaar7709} (\bibinfo {year} {2019})}\BibitemShut
	{NoStop}%
	\bibitem [{\citenamefont {Chen}\ \emph {et~al.}(2020)\citenamefont {Chen},
		\citenamefont {Liu}, \citenamefont {Luan}, \citenamefont {Liu}, \citenamefont
		{Wang}, \citenamefont {Zhu}, \citenamefont {Li}, \citenamefont {Gu},
		\citenamefont {Liang}, \citenamefont {Gao}, \citenamefont {Lu}, \citenamefont
		{Ge}, \citenamefont {Zhang}, \citenamefont {Zhu},\ and\ \citenamefont
		{Ma}}]{chen2020revealing}%
	\BibitemOpen
	\bibfield  {author} {\bibinfo {author} {\bibfnamefont {H.-Z.}\ \bibnamefont
			{Chen}}, \bibinfo {author} {\bibfnamefont {T.}~\bibnamefont {Liu}}, \bibinfo
		{author} {\bibfnamefont {H.-Y.}\ \bibnamefont {Luan}}, \bibinfo {author}
		{\bibfnamefont {R.-J.}\ \bibnamefont {Liu}}, \bibinfo {author} {\bibfnamefont
			{X.-Y.}\ \bibnamefont {Wang}}, \bibinfo {author} {\bibfnamefont {X.-F.}\
			\bibnamefont {Zhu}}, \bibinfo {author} {\bibfnamefont {Y.-B.}\ \bibnamefont
			{Li}}, \bibinfo {author} {\bibfnamefont {Z.-M.}\ \bibnamefont {Gu}}, \bibinfo
		{author} {\bibfnamefont {S.-J.}\ \bibnamefont {Liang}}, \bibinfo {author}
		{\bibfnamefont {H.}~\bibnamefont {Gao}}, \bibinfo {author} {\bibfnamefont
			{L.}~\bibnamefont {Lu}}, \bibinfo {author} {\bibfnamefont {L.}~\bibnamefont
			{Ge}}, \bibinfo {author} {\bibfnamefont {S.}~\bibnamefont {Zhang}}, \bibinfo
		{author} {\bibfnamefont {J.}~\bibnamefont {Zhu}},\ and\ \bibinfo {author}
		{\bibfnamefont {R.-M.}\ \bibnamefont {Ma}},\ }\bibfield  {title} {\bibinfo
		{title} {{Revealing the missing dimension at an exceptional point}},\ }\href
	{https://doi.org/10.1038/s41567-020-0807-y} {\bibfield  {journal} {\bibinfo
			{journal} {Nat. Phys.}\ }\textbf {\bibinfo {volume} {16}},\ \bibinfo {pages}
		{571} (\bibinfo {year} {2020})}\BibitemShut {NoStop}%
	\bibitem [{\citenamefont {Dembowski}\ \emph {et~al.}(2004)\citenamefont
		{Dembowski}, \citenamefont {Dietz}, \citenamefont {Gr{\"a}f}, \citenamefont
		{Harney}, \citenamefont {Heine}, \citenamefont {Heiss},\ and\ \citenamefont
		{Richter}}]{PhysRevE.69.056216}%
	\BibitemOpen
	\bibfield  {author} {\bibinfo {author} {\bibfnamefont {C.}~\bibnamefont
			{Dembowski}}, \bibinfo {author} {\bibfnamefont {B.}~\bibnamefont {Dietz}},
		\bibinfo {author} {\bibfnamefont {H.-D.}\ \bibnamefont {Gr{\"a}f}}, \bibinfo
		{author} {\bibfnamefont {H.~L.}\ \bibnamefont {Harney}}, \bibinfo {author}
		{\bibfnamefont {A.}~\bibnamefont {Heine}}, \bibinfo {author} {\bibfnamefont
			{W.~D.}\ \bibnamefont {Heiss}},\ and\ \bibinfo {author} {\bibfnamefont
			{A.}~\bibnamefont {Richter}},\ }\bibfield  {title} {\bibinfo {title}
		{{Encircling an exceptional point}},\ }\href
	{https://doi.org/10.1103/PhysRevE.69.056216} {\bibfield  {journal} {\bibinfo
			{journal} {Phys. Rev. E}\ }\textbf {\bibinfo {volume} {69}},\ \bibinfo
		{pages} {056216} (\bibinfo {year} {2004})}\BibitemShut {NoStop}%
	\bibitem [{\citenamefont {Doppler}\ \emph {et~al.}(2016)\citenamefont
		{Doppler}, \citenamefont {Mailybaev}, \citenamefont {B{\"o}hm}, \citenamefont
		{Kuhl}, \citenamefont {Girschik}, \citenamefont {Libisch}, \citenamefont
		{Milburn}, \citenamefont {Rabl}, \citenamefont {Moiseyev},\ and\
		\citenamefont {Rotter}}]{doppler2016dynamically}%
	\BibitemOpen
	\bibfield  {author} {\bibinfo {author} {\bibfnamefont {J.}~\bibnamefont
			{Doppler}}, \bibinfo {author} {\bibfnamefont {A.~A.}\ \bibnamefont
			{Mailybaev}}, \bibinfo {author} {\bibfnamefont {J.}~\bibnamefont {B{\"o}hm}},
		\bibinfo {author} {\bibfnamefont {U.}~\bibnamefont {Kuhl}}, \bibinfo {author}
		{\bibfnamefont {A.}~\bibnamefont {Girschik}}, \bibinfo {author}
		{\bibfnamefont {F.}~\bibnamefont {Libisch}}, \bibinfo {author} {\bibfnamefont
			{T.~J.}\ \bibnamefont {Milburn}}, \bibinfo {author} {\bibfnamefont
			{P.}~\bibnamefont {Rabl}}, \bibinfo {author} {\bibfnamefont {N.}~\bibnamefont
			{Moiseyev}},\ and\ \bibinfo {author} {\bibfnamefont {S.}~\bibnamefont
			{Rotter}},\ }\bibfield  {title} {\bibinfo {title} {{Dynamically encircling an
				exceptional point for asymmetric mode switching}},\ }\href
	{https://doi.org/10.1038/nature18605} {\bibfield  {journal} {\bibinfo
			{journal} {Nature}\ }\textbf {\bibinfo {volume} {537}},\ \bibinfo {pages}
		{76} (\bibinfo {year} {2016})}\BibitemShut {NoStop}%
	\bibitem [{\citenamefont {Su}\ \emph {et~al.}(1979)\citenamefont {Su},
		\citenamefont {Schrieffer},\ and\ \citenamefont
		{Heeger}}]{PhysRevLett.42.1698}%
	\BibitemOpen
	\bibfield  {author} {\bibinfo {author} {\bibfnamefont {W.~P.}\ \bibnamefont
			{Su}}, \bibinfo {author} {\bibfnamefont {J.~R.}\ \bibnamefont {Schrieffer}},\
		and\ \bibinfo {author} {\bibfnamefont {A.~J.}\ \bibnamefont {Heeger}},\
	}\bibfield  {title} {\bibinfo {title} {{Solitons in polyacetylene}},\ }\href
	{https://doi.org/10.1103/PhysRevLett.42.1698} {\bibfield  {journal} {\bibinfo
			{journal} {Phys. Rev. Lett.}\ }\textbf {\bibinfo {volume} {42}},\ \bibinfo
		{pages} {1698} (\bibinfo {year} {1979})}\BibitemShut {NoStop}%
	\bibitem [{\citenamefont {Longhi}(2019)}]{PhysRevResearch.1.023013}%
	\BibitemOpen
	\bibfield  {author} {\bibinfo {author} {\bibfnamefont {S.}~\bibnamefont
			{Longhi}},\ }\bibfield  {title} {\bibinfo {title} {{Probing {n}on-{H}ermitian
				skin effect and {n}on-{B}loch phase transitions}},\ }\href
	{https://doi.org/10.1103/PhysRevResearch.1.023013} {\bibfield  {journal}
		{\bibinfo  {journal} {Phys. Rev. Res.}\ }\textbf {\bibinfo {volume} {1}},\
		\bibinfo {pages} {023013} (\bibinfo {year} {2019})}\BibitemShut {NoStop}%
	\bibitem [{\citenamefont {Lin}\ and\ \citenamefont
		{Li}(2024)}]{PhysRevB.109.155137}%
	\BibitemOpen
	\bibfield  {author} {\bibinfo {author} {\bibfnamefont {R.}~\bibnamefont
			{Lin}}\ and\ \bibinfo {author} {\bibfnamefont {L.}~\bibnamefont {Li}},\
	}\bibfield  {title} {\bibinfo {title} {{Topologically compatible
				{n}on-{H}ermitian skin effect}},\ }\href
	{https://doi.org/10.1103/PhysRevB.109.155137} {\bibfield  {journal} {\bibinfo
			{journal} {Phys. Rev. B}\ }\textbf {\bibinfo {volume} {109}},\ \bibinfo
		{pages} {155137} (\bibinfo {year} {2024})}\BibitemShut {NoStop}%
	\bibitem [{\citenamefont {Xu}\ \emph {et~al.}(2020)\citenamefont {Xu},
		\citenamefont {Zhang}, \citenamefont {Chen}, \citenamefont {Fu},\ and\
		\citenamefont {Zhang}}]{PhysRevA.101.013635}%
	\BibitemOpen
	\bibfield  {author} {\bibinfo {author} {\bibfnamefont {Z.}~\bibnamefont
			{Xu}}, \bibinfo {author} {\bibfnamefont {R.}~\bibnamefont {Zhang}}, \bibinfo
		{author} {\bibfnamefont {S.}~\bibnamefont {Chen}}, \bibinfo {author}
		{\bibfnamefont {L.}~\bibnamefont {Fu}},\ and\ \bibinfo {author}
		{\bibfnamefont {Y.}~\bibnamefont {Zhang}},\ }\bibfield  {title} {\bibinfo
		{title} {{Fate of zero modes in a finite {S}u-{S}chrieffer-{H}eeger model
				with PT symmetry}},\ }\href {https://doi.org/10.1103/PhysRevA.101.013635}
	{\bibfield  {journal} {\bibinfo  {journal} {Phys. Rev. A}\ }\textbf {\bibinfo
			{volume} {101}},\ \bibinfo {pages} {013635} (\bibinfo {year}
		{2020})}\BibitemShut {NoStop}%
	\bibitem [{\citenamefont {Rafi-Ul-Islam}\ \emph {et~al.}(2022)\citenamefont
		{Rafi-Ul-Islam}, \citenamefont {Siu}, \citenamefont {Sahin}, \citenamefont
		{Lee},\ and\ \citenamefont {Jalil}}]{PhysRevResearch.4.013243}%
	\BibitemOpen
	\bibfield  {author} {\bibinfo {author} {\bibfnamefont {S.~M.}\ \bibnamefont
			{Rafi-Ul-Islam}}, \bibinfo {author} {\bibfnamefont {Z.~B.}\ \bibnamefont
			{Siu}}, \bibinfo {author} {\bibfnamefont {H.}~\bibnamefont {Sahin}}, \bibinfo
		{author} {\bibfnamefont {C.~H.}\ \bibnamefont {Lee}},\ and\ \bibinfo {author}
		{\bibfnamefont {M.~B.~A.}\ \bibnamefont {Jalil}},\ }\bibfield  {title}
	{\bibinfo {title} {{Critical hybridization of skin modes in coupled
				{n}on-{H}ermitian chains}},\ }\href
	{https://doi.org/10.1103/PhysRevResearch.4.013243} {\bibfield  {journal}
		{\bibinfo  {journal} {Phys. Rev. Res.}\ }\textbf {\bibinfo {volume} {4}},\
		\bibinfo {pages} {013243} (\bibinfo {year} {2022})}\BibitemShut {NoStop}%
	\bibitem [{\citenamefont {Rafi-Ul-Islam}\ \emph {et~al.}(2025)\citenamefont
		{Rafi-Ul-Islam}, \citenamefont {Siu}, \citenamefont {Razo},\ and\
		\citenamefont {Jalil}}]{PhysRevB.111.115415}%
	\BibitemOpen
	\bibfield  {author} {\bibinfo {author} {\bibfnamefont {S.~M.}\ \bibnamefont
			{Rafi-Ul-Islam}}, \bibinfo {author} {\bibfnamefont {Z.~B.}\ \bibnamefont
			{Siu}}, \bibinfo {author} {\bibfnamefont {M.~S.~H.}\ \bibnamefont {Razo}},\
		and\ \bibinfo {author} {\bibfnamefont {M.~B.~A.}\ \bibnamefont {Jalil}},\
	}\bibfield  {title} {\bibinfo {title} {{Critical {n}on-{H}ermitian skin
				effect in a cross-coupled {H}ermitian chain}},\ }\href
	{https://doi.org/10.1103/PhysRevB.111.115415} {\bibfield  {journal} {\bibinfo
			{journal} {Phys. Rev. B}\ }\textbf {\bibinfo {volume} {111}},\ \bibinfo
		{pages} {115415} (\bibinfo {year} {2025})}\BibitemShut {NoStop}%
	\bibitem [{\citenamefont {Li}\ \emph {et~al.}(2020)\citenamefont {Li},
		\citenamefont {Lee},\ and\ \citenamefont {Gong}}]{PhysRevLett.124.250402}%
	\BibitemOpen
	\bibfield  {author} {\bibinfo {author} {\bibfnamefont {L.}~\bibnamefont
			{Li}}, \bibinfo {author} {\bibfnamefont {C.~H.}\ \bibnamefont {Lee}},\ and\
		\bibinfo {author} {\bibfnamefont {J.}~\bibnamefont {Gong}},\ }\bibfield
	{title} {\bibinfo {title} {{Topological switch for {n}on-{H}ermitian skin
				effect in cold-atom systems with loss}},\ }\href
	{https://doi.org/10.1103/PhysRevLett.124.250402} {\bibfield  {journal}
		{\bibinfo  {journal} {Phys. Rev. Lett.}\ }\textbf {\bibinfo {volume} {124}},\
		\bibinfo {pages} {250402} (\bibinfo {year} {2020})}\BibitemShut {NoStop}%
	\bibitem [{\citenamefont {Zhao}\ \emph {et~al.}(2025)\citenamefont {Zhao},
		\citenamefont {Wang}, \citenamefont {He}, \citenamefont {Poon}, \citenamefont
		{Pak}, \citenamefont {Liu}, \citenamefont {Ren}, \citenamefont {Liu},\ and\
		\citenamefont {Jo}}]{zhao2025two}%
	\BibitemOpen
	\bibfield  {author} {\bibinfo {author} {\bibfnamefont {E.}~\bibnamefont
			{Zhao}}, \bibinfo {author} {\bibfnamefont {Z.}~\bibnamefont {Wang}}, \bibinfo
		{author} {\bibfnamefont {C.}~\bibnamefont {He}}, \bibinfo {author}
		{\bibfnamefont {T.~F.~J.}\ \bibnamefont {Poon}}, \bibinfo {author}
		{\bibfnamefont {K.~K.}\ \bibnamefont {Pak}}, \bibinfo {author} {\bibfnamefont
			{Y.-J.}\ \bibnamefont {Liu}}, \bibinfo {author} {\bibfnamefont
			{P.}~\bibnamefont {Ren}}, \bibinfo {author} {\bibfnamefont {X.-J.}\
			\bibnamefont {Liu}},\ and\ \bibinfo {author} {\bibfnamefont {G.-B.}\
			\bibnamefont {Jo}},\ }\bibfield  {title} {\bibinfo {title} {{Two-dimensional
				{n}on-{H}ermitian skin effect in an ultracold {F}ermi gas}},\ }\href
	{https://doi.org/10.1038/s41586-024-08347-3} {\bibfield  {journal} {\bibinfo
			{journal} {Nature}\ }\textbf {\bibinfo {volume} {637}},\ \bibinfo {pages}
		{565} (\bibinfo {year} {2025})}\BibitemShut {NoStop}%
	\bibitem [{\citenamefont {Wang}\ \emph {et~al.}(2025)\citenamefont {Wang},
		\citenamefont {Xu},\ and\ \citenamefont {Li}}]{wang2025topological}%
	\BibitemOpen
	\bibfield  {author} {\bibinfo {author} {\bibfnamefont {H.}~\bibnamefont
			{Wang}}, \bibinfo {author} {\bibfnamefont {Z.}~\bibnamefont {Xu}},\ and\
		\bibinfo {author} {\bibfnamefont {Z.}~\bibnamefont {Li}},\ }\bibfield
	{title} {\bibinfo {title} {{Topological phase transitions and edge-state
				transfer in time-multiplexed quantum walks}},\ }\href
	{https://doi.org/10.1103/sb89-p5rl} {\bibfield  {journal} {\bibinfo
			{journal} {Phys. Rev. A}\ }\textbf {\bibinfo {volume} {112}},\ \bibinfo
		{pages} {042230} (\bibinfo {year} {2025})}\BibitemShut {NoStop}%
	\bibitem [{\citenamefont {Li}\ \emph {et~al.}(2021)\citenamefont {Li},
		\citenamefont {Lee},\ and\ \citenamefont {Gong}}]{li2021impurity}%
	\BibitemOpen
	\bibfield  {author} {\bibinfo {author} {\bibfnamefont {L.}~\bibnamefont
			{Li}}, \bibinfo {author} {\bibfnamefont {C.~H.}\ \bibnamefont {Lee}},\ and\
		\bibinfo {author} {\bibfnamefont {J.}~\bibnamefont {Gong}},\ }\bibfield
	{title} {\bibinfo {title} {{Impurity induced scale-free localization}},\
	}\href {https://doi.org/10.1038/s42005-021-00547-x} {\bibfield  {journal}
		{\bibinfo  {journal} {Commun. Phys.}\ }\textbf {\bibinfo {volume} {4}},\
		\bibinfo {pages} {42} (\bibinfo {year} {2021})}\BibitemShut {NoStop}%
	\bibitem [{\citenamefont {Liu}\ and\ \citenamefont
		{Xu}(2023)}]{PhysRevB.108.184205}%
	\BibitemOpen
	\bibfield  {author} {\bibinfo {author} {\bibfnamefont {J.}~\bibnamefont
			{Liu}}\ and\ \bibinfo {author} {\bibfnamefont {Z.}~\bibnamefont {Xu}},\
	}\bibfield  {title} {\bibinfo {title} {{From ergodicity to many-body
				localization in a one-dimensional interacting {n}on-{H}ermitian {S}tark
				system}},\ }\href {https://doi.org/10.1103/PhysRevB.108.184205} {\bibfield
		{journal} {\bibinfo  {journal} {Phys. Rev. B}\ }\textbf {\bibinfo {volume}
			{108}},\ \bibinfo {pages} {184205} (\bibinfo {year} {2023})}\BibitemShut
	{NoStop}%
	\bibitem [{\citenamefont {Xu}\ and\ \citenamefont
		{Chen}(2020)}]{PhysRevB.102.035153}%
	\BibitemOpen
	\bibfield  {author} {\bibinfo {author} {\bibfnamefont {Z.}~\bibnamefont
			{Xu}}\ and\ \bibinfo {author} {\bibfnamefont {S.}~\bibnamefont {Chen}},\
	}\bibfield  {title} {\bibinfo {title} {{Topological {B}ose-{M}ott insulators
				in one-dimensional {n}on-{H}ermitian superlattices}},\ }\href
	{https://doi.org/10.1103/PhysRevB.102.035153} {\bibfield  {journal} {\bibinfo
			{journal} {Phys. Rev. B}\ }\textbf {\bibinfo {volume} {102}},\ \bibinfo
		{pages} {035153} (\bibinfo {year} {2020})}\BibitemShut {NoStop}%
	\bibitem [{\citenamefont {Qin}\ \emph {et~al.}(2026)\citenamefont {Qin},
		\citenamefont {Ang}, \citenamefont {Lee},\ and\ \citenamefont
		{Li}}]{qin2026many}%
	\BibitemOpen
	\bibfield  {author} {\bibinfo {author} {\bibfnamefont {Y.}~\bibnamefont
			{Qin}}, \bibinfo {author} {\bibfnamefont {Y.~S.}\ \bibnamefont {Ang}},
		\bibinfo {author} {\bibfnamefont {C.~H.}\ \bibnamefont {Lee}},\ and\ \bibinfo
		{author} {\bibfnamefont {L.}~\bibnamefont {Li}},\ }\bibfield  {title}
	{\bibinfo {title} {{Many-body critical {n}on-{H}ermitian skin effect}},\
	}\href {https://doi.org/10.1038/s42005-025-02448-9} {\bibfield  {journal}
		{\bibinfo  {journal} {Commun. Phys.}\ }\textbf {\bibinfo {volume} {9}},\
		\bibinfo {pages} {16} (\bibinfo {year} {2026})}\BibitemShut {NoStop}%
	\bibitem [{\citenamefont {Liu}\ \emph {et~al.}(2025)\citenamefont {Liu},
		\citenamefont {Jiang}, \citenamefont {Xue}, \citenamefont {Li}, \citenamefont
		{Gong}, \citenamefont {Liu},\ and\ \citenamefont
		{Lee}}]{liu2025entanglement}%
	\BibitemOpen
	\bibfield  {author} {\bibinfo {author} {\bibfnamefont {S.}~\bibnamefont
			{Liu}}, \bibinfo {author} {\bibfnamefont {H.}~\bibnamefont {Jiang}}, \bibinfo
		{author} {\bibfnamefont {W.-T.}\ \bibnamefont {Xue}}, \bibinfo {author}
		{\bibfnamefont {Q.}~\bibnamefont {Li}}, \bibinfo {author} {\bibfnamefont
			{J.}~\bibnamefont {Gong}}, \bibinfo {author} {\bibfnamefont {X.}~\bibnamefont
			{Liu}},\ and\ \bibinfo {author} {\bibfnamefont {C.~H.}\ \bibnamefont {Lee}},\
	}\bibfield  {title} {\bibinfo {title} {{{N}on-{H}ermitian entanglement dip
				from scaling-induced exceptional criticality}},\ }\href
	{https://doi.org/10.1016/j.scib.2025.07.011} {\bibfield  {journal} {\bibinfo
			{journal} {Sci. Bull.}\ }\textbf {\bibinfo {volume} {70}},\ \bibinfo {pages}
		{2929} (\bibinfo {year} {2025})}\BibitemShut {NoStop}%
	\bibitem [{\citenamefont {Yang}\ and\ \citenamefont
		{Lee}(2024)}]{PhysRevLett.133.136602}%
	\BibitemOpen
	\bibfield  {author} {\bibinfo {author} {\bibfnamefont {M.}~\bibnamefont
			{Yang}}\ and\ \bibinfo {author} {\bibfnamefont {C.~H.}\ \bibnamefont {Lee}},\
	}\bibfield  {title} {\bibinfo {title} {{Percolation-induced PT symmetry
				breaking}},\ }\href {https://doi.org/10.1103/PhysRevLett.133.136602}
	{\bibfield  {journal} {\bibinfo  {journal} {Phys. Rev. Lett.}\ }\textbf
		{\bibinfo {volume} {133}},\ \bibinfo {pages} {136602} (\bibinfo {year}
		{2024})}\BibitemShut {NoStop}%
	\bibitem [{\citenamefont {Wang}\ \emph {et~al.}(2021)\citenamefont {Wang},
		\citenamefont {Dutt}, \citenamefont {Yang}, \citenamefont {Wojcik},
		\citenamefont {Vu{\v{c}}kovi{\'c}},\ and\ \citenamefont
		{Fan}}]{wang2021generating}%
	\BibitemOpen
	\bibfield  {author} {\bibinfo {author} {\bibfnamefont {K.}~\bibnamefont
			{Wang}}, \bibinfo {author} {\bibfnamefont {A.}~\bibnamefont {Dutt}}, \bibinfo
		{author} {\bibfnamefont {K.~Y.}\ \bibnamefont {Yang}}, \bibinfo {author}
		{\bibfnamefont {C.~C.}\ \bibnamefont {Wojcik}}, \bibinfo {author}
		{\bibfnamefont {J.}~\bibnamefont {Vu{\v{c}}kovi{\'c}}},\ and\ \bibinfo
		{author} {\bibfnamefont {S.}~\bibnamefont {Fan}},\ }\bibfield  {title}
	{\bibinfo {title} {{Generating arbitrary topological windings of a
				{n}on-{H}ermitian band}},\ }\href {https://doi.org/10.1126/science.abf6568}
	{\bibfield  {journal} {\bibinfo  {journal} {Science}\ }\textbf {\bibinfo
			{volume} {371}},\ \bibinfo {pages} {1240} (\bibinfo {year}
		{2021})}\BibitemShut {NoStop}%
	\bibitem [{\citenamefont {Park}\ \emph {et~al.}(2022)\citenamefont {Park},
		\citenamefont {Cho}, \citenamefont {Lee}, \citenamefont {Lee}, \citenamefont
		{Lee}, \citenamefont {Park}, \citenamefont {Ryu}, \citenamefont {Park},
		\citenamefont {Jeon},\ and\ \citenamefont {Min}}]{park2022revealing}%
	\BibitemOpen
	\bibfield  {author} {\bibinfo {author} {\bibfnamefont {J.}~\bibnamefont
			{Park}}, \bibinfo {author} {\bibfnamefont {H.}~\bibnamefont {Cho}}, \bibinfo
		{author} {\bibfnamefont {S.}~\bibnamefont {Lee}}, \bibinfo {author}
		{\bibfnamefont {K.}~\bibnamefont {Lee}}, \bibinfo {author} {\bibfnamefont
			{K.}~\bibnamefont {Lee}}, \bibinfo {author} {\bibfnamefont {H.~C.}\
			\bibnamefont {Park}}, \bibinfo {author} {\bibfnamefont {J.-W.}\ \bibnamefont
			{Ryu}}, \bibinfo {author} {\bibfnamefont {N.}~\bibnamefont {Park}}, \bibinfo
		{author} {\bibfnamefont {S.}~\bibnamefont {Jeon}},\ and\ \bibinfo {author}
		{\bibfnamefont {B.}~\bibnamefont {Min}},\ }\bibfield  {title} {\bibinfo
		{title} {{Revealing {n}on-{H}ermitian band structure of photonic {F}loquet
				media}},\ }\href {https://doi.org/10.1126/sciadv.abo6220} {\bibfield
		{journal} {\bibinfo  {journal} {Sci. Adv.}\ }\textbf {\bibinfo {volume}
			{8}},\ \bibinfo {pages} {eabo6220} (\bibinfo {year} {2022})}\BibitemShut
	{NoStop}%
	\bibitem [{\citenamefont {Cao}\ \emph {et~al.}(2023)\citenamefont {Cao},
		\citenamefont {Li}, \citenamefont {Zhao}, \citenamefont {Guo}, \citenamefont
		{Qi}, \citenamefont {Chang}, \citenamefont {Zhou}, \citenamefont {Xu},\ and\
		\citenamefont {Duan}}]{PhysRevLett.130.163001}%
	\BibitemOpen
	\bibfield  {author} {\bibinfo {author} {\bibfnamefont {M.-M.}\ \bibnamefont
			{Cao}}, \bibinfo {author} {\bibfnamefont {K.}~\bibnamefont {Li}}, \bibinfo
		{author} {\bibfnamefont {W.-D.}\ \bibnamefont {Zhao}}, \bibinfo {author}
		{\bibfnamefont {W.-X.}\ \bibnamefont {Guo}}, \bibinfo {author} {\bibfnamefont
			{B.-X.}\ \bibnamefont {Qi}}, \bibinfo {author} {\bibfnamefont {X.-Y.}\
			\bibnamefont {Chang}}, \bibinfo {author} {\bibfnamefont {Z.-C.}\ \bibnamefont
			{Zhou}}, \bibinfo {author} {\bibfnamefont {Y.}~\bibnamefont {Xu}},\ and\
		\bibinfo {author} {\bibfnamefont {L.-M.}\ \bibnamefont {Duan}},\ }\bibfield
	{title} {\bibinfo {title} {{Probing complex-energy topology via
				{n}on-{H}ermitian absorption spectroscopy in a trapped ion simulator}},\
	}\href {https://doi.org/10.1103/PhysRevLett.130.163001} {\bibfield  {journal}
		{\bibinfo  {journal} {Phys. Rev. Lett.}\ }\textbf {\bibinfo {volume} {130}},\
		\bibinfo {pages} {163001} (\bibinfo {year} {2023})}\BibitemShut {NoStop}%
\end{thebibliography}
\end{document}